\documentclass[11pt, draftclsnofoot, onecolumn]{IEEEtran}

\usepackage[pdftex]{graphicx}
\graphicspath{{figures/}}
\usepackage[para,online,flushleft]{threeparttable}

\usepackage{amssymb,amsmath}
\usepackage[dvipsnames]{xcolor}
\usepackage{url}
\usepackage{comment}

\newcommand\minisec[1]{\vspace{0.8mm}\noindent\textbf{#1 ---}}

\begin{document}

\title{Audio-based Musical Version Identification:\\Elements and Challenges}
%
%
%

\author{Furkan~Yesiler,
        Guillaume~Doras,
        Rachel~M.~Bittner,
        Christopher~J.~Tralie,
        Joan~Serr\`a}

%

\markboth{IEEE Signal Processing Magazine,~Vol.~XX, No.~XX, 2021}%
{}

\maketitle



\IEEEpeerreviewmaketitle

\section{Introduction}\label{sec:intro}
Creating novel interpretations of existing musical compositions is and has always been an essential part of musical practice. Before the advent of recorded music, listening to a piece of music mostly meant listening to a version of it, in many cases, performed by musicians other than the original composer or performer.
While this practice, along with musical notation, allows compositions to remain known for many decades or even centuries, it also provides room for artistic expression. Oftentimes, versions that are reproduced faithfully with respect to the original composition are seen as tributes to honor the composers; however, versions that are altered with the limitless creativity of humans often demonstrate how an existing idea can be transformed into something that goes beyond the original intention. Regardless of the ways in which musical versions are created, they are fundamental to the world’s musical heritage.

\begin{figure}[tb]
\centering
\includegraphics[width=\columnwidth]{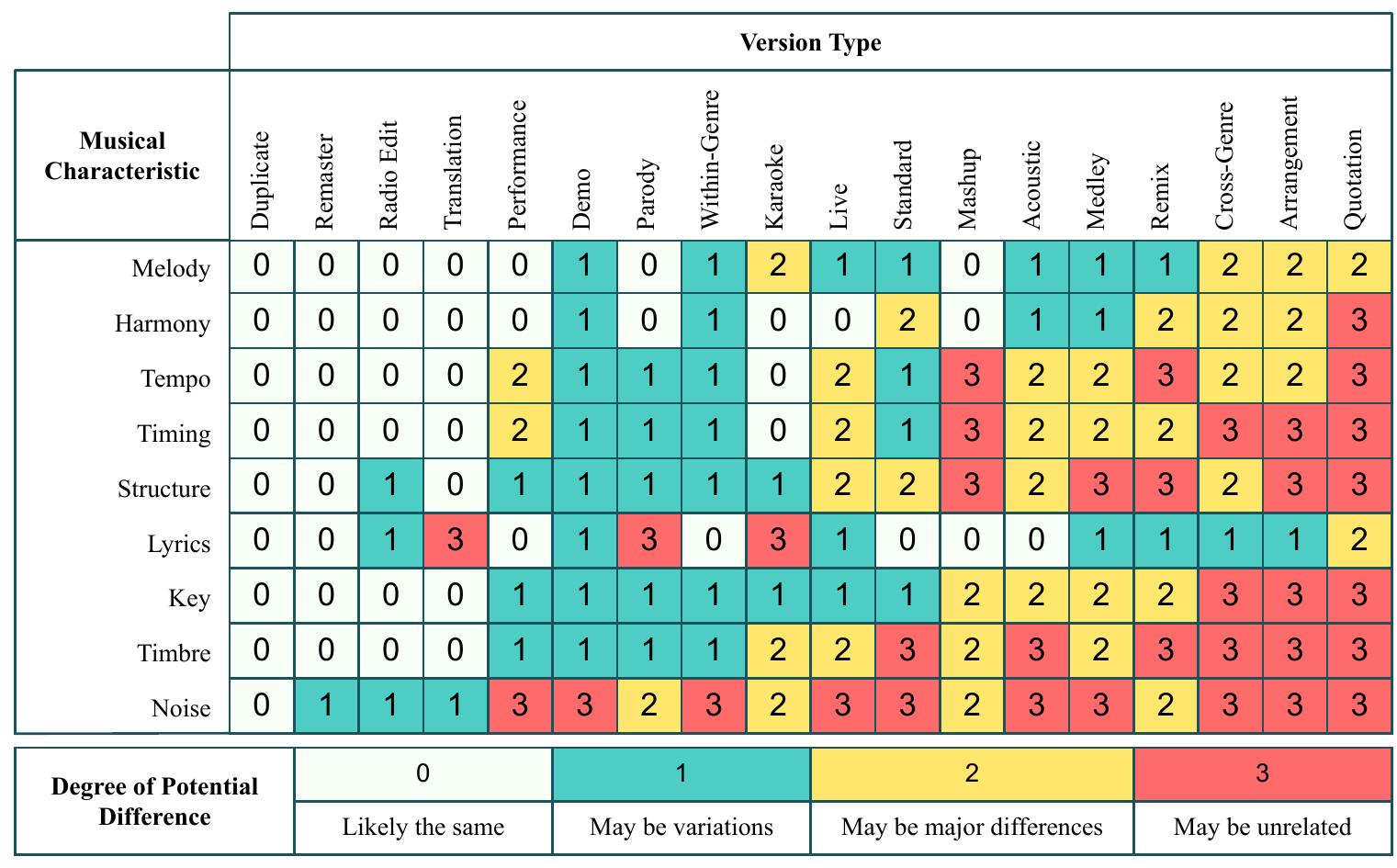}
\caption{Examples of version types and the (subjective) degrees to which a list of musical dimensions may be altered. Both the version types and the musical dimensions are based on~\cite{serra2010audio} with a few additions. The version type ``Performance'' refers to a recording of a written classical work, while ``Standard'' refers to a recording of a folk or a jazz tune where there are often improvisational aspects.}
\label{fig:version_types}
\end{figure}

In this article, we consider a musical version to be ``any rendition or performance of 
an existing musical composition.''
Another widely used term in the literature is ``cover songs''; however, there are three main reasons we avoid using that term in this article. 
Firstly, the scope of the term is not clear as many authors limit its use to certain genres (e.g., pop and rock) and time periods (e.g.,~after the 1950s). 
Secondly, it has negative historical and economic connotations that date back to when the term emerged: record labels often asked White musicians to record (or cover) the songs of Black musicians as a marketing strategy to reach certain demographics. 
Lastly, the term does not represent the wide range of musical versions that are the subject of this article (see Fig.~\ref{fig:version_types}). 
Therefore, we employ the term ``musical versions'' throughout this article for being as inclusive as possible.

This survey article focuses on the literature studying musical version identification and retrieval from a computational perspective.
A key concept we refer to throughout this article is the concept of similarity, which we aim to quantify and model in a way that would reflect musical versions as semantically closer than non-versions.
Following the literature, we consider that all versions of the same composition belong to the same group, or clique.

Studies in audio-based music information retrieval (MIR) focus on extracting information from audio signals (tracks), which is then exploited to develop technologies that can be used for various applications including music retrieval, recommendation, and classification. 
Following a query-by-example paradigm, such applications require a notion of musical similarity. However, considering the complexity of information carried by musical audio signals, defining a single notion of similarity is a rather difficult and perhaps futile goal. 
Therefore, the scope of musical similarity 
for various MIR tasks 
can be situated on a two-dimensional plane characterized by specificity and granularity~\cite{grosche2012audio}. The two ends of the specificity axis contain high- and low-specificity systems that are differentiated by the degree of similarity between their queries and targets. While high-specificity systems aim to identify the exact musical tracks (e.g.,~music fingerprinting), low-specificity systems are concerned with broader descriptions of music (e.g.,~genre, mood, and instruments) to retrieve tracks that are related to a given query from high-level musical properties. In terms of granularity, MIR tasks are situated on a spectrum that goes from fragment- to document-level retrieval scenarios. In fragment-level scenarios, queries and targets are short fragments of audio tracks while in document-level scenarios, they are mostly entire audio tracks.

Defining the concept of similarity that connects musical versions is a challenging task. As humans, we can, in most cases, easily identify two tracks as versions of one another. However, considering the wide range of version types~\cite{serra2010audio} (see Fig.~\ref{fig:version_types}), constructing a comprehensive similarity definition is extremely difficult. Such versions may incorporate various differences in musical dimensions~\cite{serra2010audio}, including differences in timbre, tempo, structure, lyrics, recording conditions (``noise''), and so on (Fig.~\ref{fig:version_types}).
For example, a ``radio edit'' of a track may have minor differences in recording quality, have sections removed, and have non-explicit lyrics, but all other musical dimensions remain mostly unchanged.
Live versions of a track often have higher degrees of variation: they may have small differences in the melody, key, and lyrics; more drastic variations in tempo, timing, structure, and timbre/instrumentation; and lots of background noise from the live recording environment.
Remixes, on the other hand, may have very little in common with the original, sharing only lyrics and melody for example, which may be superimposed on 
musical content from a different track.

Due to such differences, the connections that link musical versions together vary depending on each case. For instance, while some version pairs may share the same melodic phrases, 
others may share only the lyrics. Therefore, modeling the information shared by various types of versions requires a similarity notion that incorporates multiple musical dimensions. Formulating such a notion from a computational perspective is the main focus of the line of research often referred to as version identification (VI).
Note that, in this article, we focus on techniques that address a wide variety of versions simultaneously. However, there are a number of subfields (discussed in Sec.~\ref{sec:sub-fields}) that are built specifically for particular types of versions.
On the aforementioned specificity--granularity plane, VI can be situated as a task that is mid-specificity and document-level, as the degree of similarity is neither based on high-level concepts nor exact characteristics of signals and the queries and targets are often entire tracks.

The first efforts toward VI emerged in the early 2000s~\cite{foote2000arthur}, and it has remained an active field of research ever since. VI systems are designed in a query-by-example fashion: given a query, the goal is to retrieve all the different interpretations of the same musical composition from a corpus. 
The main consideration for building a VI system is to overlook the differences in musical characteristics and focus on the 
shared information connecting version pairs.
However, instead of aiming to directly quantify 
this shared information, such
systems create representations that are invariant to the aforementioned differences. In light of this, VI research, as other music retrieval~\cite{muller2019} and classification~\cite{nam2019} studies in MIR, benefits from the advancements from many scientific disciplines such as signal processing, machine learning, non-linear time series analysis, computational biology, etc.

The potential impact of VI research can be examined from three perspectives. Firstly, from an MIR perspective, it holds the promise of better quantifying (leading ultimately to better understanding of) the connections between musical versions.
Previous research focuses on using the harmonic and the melodic characteristics of musical audio to formulate the concept of similarity for this task. Although providing plausible results, solely relying on these two musical characteristics is not sufficient to cover all the variety that can be found across versions. Therefore, exploring novel ideas that exploit a more comprehensive definition of similarity in the VI context is still an ongoing challenge. Moreover, from a more musicological aspect, understanding how versions are connected and how they evolve through time may reveal valuable lessons for better analysis of artistic inspirations in the music creation process.
Secondly, from an industrial perspective, building accurate and scalable VI systems benefits various needs in the current music ecosystem, including, but not limited to, detection of copyright infringements in media platforms (e.g., YouTube, Apple, and Spotify) and for author and composer societies (e.g., SACEM\footnote{\url{https://www.sacem.fr/en}}, SGAE\footnote{\url{http://www.sgae.es/en-EN/SitePages/index.aspx}}, and ASCAP\footnote{\url{https://www.ascap.com/}}), organizing and navigating through vast music corpora, and automatically identifying versions in live performances. 
Due to the 
rapid
increase in the amount of new musical content created and uploaded to media platforms, automating such application scenarios is becoming increasingly important.
Lastly, from a
listener
perspective, finding new interpretations of a favorite song may have value for the listening experience and for music appreciation. 
The existence of many websites\footnote{\url{https://secondhandsongs.com/}}\textsuperscript{,}\footnote{\url{https://www.whosampled.com/}} that are dedicated to annotating and sharing this information can be seen as an indicator of user interest in versions in general.

In this article, we aim to provide a review of the key ideas and approaches proposed in 20 years of scientific literature around VI research and connect them to current practice. For more than a decade, VI systems suffered from the accuracy--scalability trade-off, with attempts to increase accuracy that typically resulted in cumbersome, non-scalable systems. Recent years, however, have witnessed the rise of deep learning--based approaches that take a step toward bridging the accuracy--scalability gap, yielding systems that can realistically be deployed in industrial applications. Although this trend positively influences the number of researchers and institutions working on VI, it may also result in obscuring the literature before the deep learning era. To appreciate two decades of novel ideas in VI research and to facilitate building better systems, we now review some of the successful concepts and applications proposed in the literature and study their evolution throughout the years. To facilitate understanding some of the concepts mentioned in this article, we include audio examples on our supplementary website\footnote{\url{https://furkanyesiler.github.io/musical_version_id_spm/}}.

The rest of the article is organized as follows. We give a chronological survey of the evolution of VI systems in Section~\ref{sec:survey}. We introduce the building blocks of VI systems and describe them in detail in Section~\ref{sec:blocks}. We then introduce a set of ideas that can be used in any VI system regardless of their building blocks in Section~\ref{sec:beyond_blocks}. We overview the publicly available datasets and evaluation metrics for VI in Section~\ref{sec:data_and_eval}. Finally, in Section~\ref{sec:issues}, we present some current open issues to provide ideas for researchers interested in working on VI.
\section{A Historical Survey of VI Systems}\label{sec:survey}
\begin{figure}[tb]
\centering
\includegraphics[width=\columnwidth]{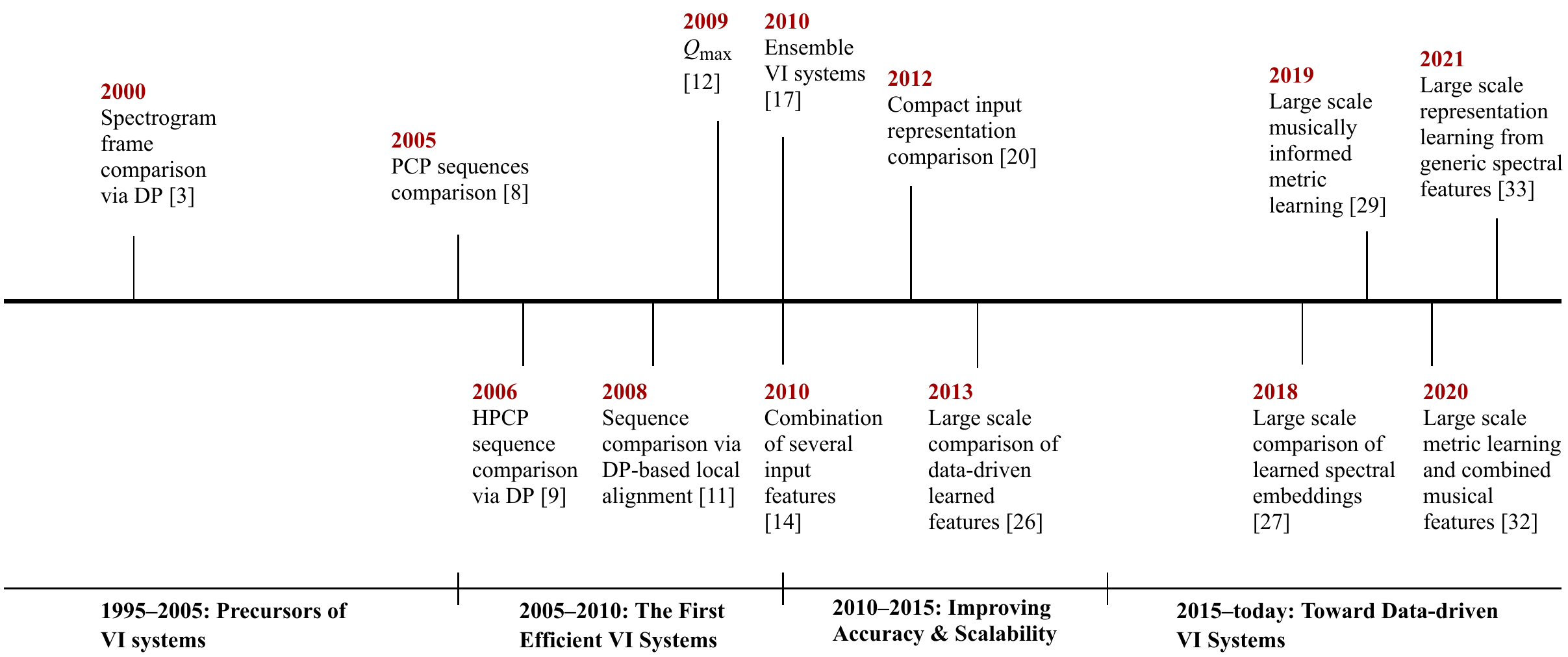}
\caption{Milestones of VI research over the past 20~years.}
\label{fig:vi_history}
\end{figure}

This section aims to provide a survey of the evolution of VI systems across 20~years of research (see Fig.~\ref{fig:vi_history} for a timeline overview of important milestones). 
We discuss how pioneering VI approaches were inspired by earlier music retrieval systems and how they were continuously improved over time to address the complex VI use case.
Throughout this section, various system components will be referenced which are later explained in Sections~\ref{sec:blocks} and~\ref{sec:beyond_blocks}. 

\subsection{1995--2005: Precursors of VI Systems}
\label{sec:music_retrieval}

Depending on the context, the musical similarity is typically assessed from editorial or social metadata such as tags or genre, from symbolic data such as Musical Instrument Digital Interface (MIDI)\footnote{https://en.wikipedia.org/wiki/MIDI} representations, or from the audio content itself such as waveform or spectral representations. Considering such kinds of similarity assessments, pioneering works of VI mainly relied on symbolic and audio data.

\minisec{Comparing discrete sequences} In the mid-1990s, pioneering music retrieval systems originally relied on symbolic musical representations, for instance, MIDI. In this formalism, a musical excerpt was described as a series of symbols: for instance, a monophonic melody could be described as a sequence of n-grams of intervals between consecutive notes, and the same principle was extended to polyphonic lines, encoding relative pitch values and durations in the n-gram. The similarity between musical excerpts, each represented as a series of symbols, was then evaluated with standard text-based comparison methods, such as regular expressions.
However, symbolic representations only exist for very specific corpora, while audio content is now commonly available and often the main source of musical creation.

The first attempts to establish musical similarity directly from audio content aimed at reducing the problem to the already known symbolic case. For instance, for query-by-humming applications,
a short hummed or whistled audio input was processed with a pitch tracker, and intervals between consecutive pitches were encoded into a series of symbols and used to query a corpus of musical scores encoded in the same way. The more complex case of polyphonic audio content was also reduced to the monophonic symbolic case by extracting a sequence of the most salient pitches.

The conversion of salient pitches into sequences of symbols ultimately relied on the limited efficiency of the then-available pitch estimation algorithms. It quickly appeared that they were not accurate enough for strict string matching, which motivated the introduction of sequence comparison algorithms based on dynamic programming (DP, see Section~\ref{sec:tempotime}). The principle was to estimate the similarity between two sequences by counting the symbol insertions or deletions that are required to align them. This method was well-suited for musical sequence comparison because symbol (e.g.,~note) insertion and deletion could be penalized based on musical plausibility (e.g.,~subsequent notes could be considered more likely to be within a small interval, thus large intervals could be more heavily penalized). Text-based comparison methods were abandoned in favor of DP comparisons, and these became the \textit{de facto} standard for musical sequence comparison (see~\cite{doras2020phd} for a summary).

However, symbolic representations are inherently discrete and turned out not to be expressive enough to embed all the musical complexity of real audio content. This fostered the development of alternative, real-valued representations. 

\minisec{Comparing real-valued sequences} One of the first attempts at using real-valued representations for VI purposes was proposed by Foote et al.~\cite{foote2000arthur}. The idea was to represent an audio excerpt by its energy profile (the root-mean-square signal power over short windows) and compare the resulting real-valued sequences via DP.

In a more musically motivated approach, another idea was to model
music as a stochastic process, in particular, as a Markov model where each state corresponds to a chord. An audio excerpt was then represented as a Markov chain, and its similarity with others was assessed with an appropriate metric, such as the Kullback-Leibler divergence. 
To estimate chords more accurately, Bello et al.~\cite{bello2005robust} proposed training a hidden Markov model based on a simple chord vocabulary using an initial beat-synchronous pitch class profile (PCP, also known as ``chroma'') sequence (see Section~\ref{sec:input_repr}). The idea was to use this model to infer the most probable chord sequence that could have generated an observed PCP sequence and to use such a chord sequence as a proxy to evaluate musical similarity.

It then appeared that the PCP sequence itself was particularly well-suited for musical comparison: it could be deterministically computed directly from the audio and adequately represented the relative intensity of each pitch class of the equal-tempered scale. This idea was used by M\"uller et al.~\cite{muller2005audio}, who proposed assessing the similarity between audio excerpts by comparing PCP features in a frame-wise manner, achieving tempo invariance using several representations of the same excerpt computed at different sampling rates.

Although not directly related as they focus on the retrieval of the music recordings that match the query in an exact manner, music fingerprinting algorithms (see~\cite{grosche2012audio}) were proposed around the same time, which hash connections between characteristic audio ``landmarks'' (i.e.,~patterns of large spectral peaks).
These algorithms were however not invariant to tempo, timbre, or pitch changes. Although they were extremely efficient for exact matching throughout the entire duration of a track, they were not designed to match different versions and usually not used for VI purposes.

\subsection{2005--2010: The First Efficient VI Systems}
\label{sec:cover_detection}

In the mid-2000s, the topic of musical similarity was being studied from many dimensions: from music fingerprinting to classifying musical genres. At the same time, VI
started to gain more attention from the community. Pioneering attempts were logically built upon existing music retrieval approaches: for instance, using pitch trackers to extract 
dominant melodies (see Section~\ref{sec:input_repr}) as a discrete input representation was one of the first proposals for VI. However, the complexity of the task pushed researchers toward exploring solutions that were better suited for this particular problem.

\minisec{Comparing harmonic features} Most of the first robust VI methods followed the same principle: they used a harmonic progression (typically a sequence of enhanced PCP), transposed it to ensure key invariance (see Section~\ref{sec:transposition_invariance}), and computed a similarity score between pairs of those resulting sequences. 
For instance, G\'omez et al.~\cite{gomez2006automatic} proposed the use of an enhanced PCP-based representation, the harmonic pitch class profile (HPCP, see Section~\ref{sec:input_repr}). Key invariance was achieved by normalizing HPCP with its estimated global key, and similarity was assessed via DP. The same year, an approach involving beat-synchronous PCP representations (see Section~\ref{sec:tempotime}) was proposed and later elaborated by Ellis and Poliner~\cite{ellis2007identifyingcover}. 
Key invariance was achieved using each possible relative PCP rotation, and similarity was assessed via cross-correlation between transposed sequences. 
This method yielded the best performance on the first Music Information Retrieval Evaluation eXchange\footnote{https://www.music-ir.org/mirex} (MIREX) ``audio cover song identification'' (i.e.,~VI) contest that took place in 2006. 
These approaches were improved further for the following 2007 and 2008 MIREX editions. 
For instance, Serr\`a et al.~\cite{serra2008chroma} used another method to ensure key invariance, named optimal transposition index (OTI, see Section~\ref{sec:transposition_invariance}), which transposed one track relative to the other so that they share the same common global key.

\minisec{Improving dynamic programming} In the same work, Serr\`a et al.~\cite{serra2008chroma} also introduced DP-based local alignment with musically motivated constraints to account for tempo and structure differences between versions, obtaining state-of-the-art results in 2007 and 2008.
The following year, Serr\`a et al.~\cite{serra2009cross} adapted several concepts commonly used to study recurrences in physical or biological systems. 
The idea was to compare PCP sequences using a cross recurrence plot (CRP), a representation highlighting common subsequences. 
The global similarity was assessed via recurrence quantification analysis measurement, which in essence quantifies the importance of the similarity patterns in the CRP. 
This algorithm was enhanced further using similarity transitivity: if A and B are similar and B and C are similar, then it is likely that A and C are similar too~\cite{serra2009unsupervised} (see Section~\ref{sec:clique_enhancement}). 
The combination of this algorithm with the previous approach~\cite{serra2009cross}, dubbed $Q^*_{\text{max}}$, remained the state-of-the-art method in VI for more than a decade.

\subsection{2010--2015: Improving Accuracy \& Scalability}

By the early 2010s, successful VI systems based on harmonic representations and local alignment had achieved promising accuracy. However, it appeared that harmonic information was not the only musical facet that could be used to adequately model music complexity. At the same time, DP algorithms were too computationally expensive to address industrial applications and the ever-increasing size of modern music corpora.

\minisec{Improving accuracy} A strategy for improving accuracy was to investigate alternative input features: for instance, some representations were designed to account for the characteristics of the human auditory perception system (see~\cite{doras2020phd} for a review).

Another strategy was to consider the combination of existing features: the idea was that different features embed complementary information, and combining them would improve the accuracy compared to using each input feature separately (see Section~\ref{sec:feature_fusion}). Several combinations were investigated: for instance, HPCP extracted from the original mixture, the separated vocals, and the separated accompaniment~\cite{foucard2010multimodal}; dominant melody, bass line, and the harmony (i.e.,~HPCP)~\cite{salamon2012melodybass}; or timbral features and HPCP~\cite{tralie2017cover}. It was shown that these combined representations improved the accuracy compared to cases where each feature was considered alone.

A third strategy was derived from the observation that some methods performed better for certain tracks than others and that an ensemble system could blend existing systems' strengths (see Section~\ref{sec:ensemble_systems}). Following this line of thought, different approaches based on classifiers~\cite{ravuri2010cover}, rank aggregation~\cite{osmalsky2016enhancing}, and similarity network fusion~\cite{tralie2017cover, chen2018} were investigated to merge scores obtained from various systems. In all cases, ensemble systems improved upon the accuracy of single systems.

\minisec{Improving scalability} All successful VI algorithms described so far rely on variants of DP sequence comparison which scale quadratically with the length of the sequences.
This time complexity quickly becomes prohibitive when querying large corpora; the sequence comparison computation with the query track must be done on the fly for every track in the corpus.
The problem of scalability is common to all information retrieval systems: in order to scale, they generally require a very lightweight similarity estimation function (e.g.,~a simple Euclidean distance), which in turn implies a data representation that can be computed offline and conveniently stored for fast lookup (e.g.,~a small vector of real numbers).

A first direction to improve scalability was to transform input features into a more compact representation, ideally a lightweight matrix (or vector) that could be compared via Frobenius norm (or Euclidean distance). A popular compacting transformation was the 2D~Fourier transform of the PCP sequences (see Section~\ref{sec:transposition_invariance}) because it provides key and time-shift invariance with respect to
the original input~\cite{bertin2012large}.

Another approach to reducing the size of input features was based on the assumptions that similarity between versions mainly depends on certain parts of the audio and that comparing short segments should be more efficient than comparing an entire track (see Section~\ref{sec:struct}). Different methods were investigated, for instance, detecting and isolating only the segments of interest, such as those exhibiting some degree of repetition~\cite{silva2015music}.

A second direction was directly inspired by fast indexing and retrieval methods that proved their efficiency in text-based retrieval contexts: instead of comparing audio features, the idea was to devise a hashing function and to store audio hashes in an index (see Section~\ref{sec:fast_indexing}). For instance, locality-sensitive hashing (LSH) was used to obtain the same hash for similar audio shingles, allowing an extremely fast lookup of musically similar audio excerpts~\cite{casey2006song}. 
Along the same vein, inverted file indexing was also adapted to the music retrieval context. 
The original idea was to index text documents by the keywords they contain. In a musical context, a codebook of audio-based tags plays the role of the keywords. These tags were obtained by vector clustering~\cite{kurth2008efficient} or seeding (indexing of fixed-length short regions)~\cite{martin2012blast}. 
 
Both lightweight input features and fast indexing yielded efficient lookup times (as fast as a few seconds on a one million track corpus) but exhibited poor accuracy compared to previous systems.

Finally, a third direction was to combine different methods in a two-step pruning approach: the first step aimed to discard tracks from the corpus that could be easily distinguished as non-versions using scalable systems~\cite{cai2016} or ``weak rejectors''~\cite{osmalsky2016enhancing}, while the second step involved a more accurate method operating on the remaining candidates (see Section~\ref{sec:pruning}).

\subsection{2015--today: Toward Data-driven VI Systems}
\label{sec:trends}

In the mid-2010s, the VI community was confronted with a dilemma: 
accurate systems, which could not scale up to industry-sized corpora, versus fast systems, which struggled in accuracy and were not suitable for practical use.

Many other MIR applications during this period were also confronted with a plateau in their performance gains. Inspired by advances made in other fields (e.g.,~computer vision, natural language processing, and speech processing), the community initiated a paradigm shift from ad-hoc hand-crafted feature extraction toward data-driven feature learning. 
This new perspective created a new opportunity for VI: to build more expressive representations of the audio, while still enabling a faster similarity estimation function.

\minisec{The use of feature learning} The representation learning paradigm aims at identifying and disentangling the underlying structures in the original data that can explain its relevant characteristics. 
Various attempts to learn a mid-level representation from the PCP-based descriptors were previously proposed, for instance with Markov models or $k$-means. 
Humphrey et al.~\cite{humphrey2013data} were however the first to explicitly propose using data-driven learned features to represent a track as a single embedding vector and estimate similarities between tracks by computing Euclidean distance between such embeddings.
Their method used $k$-means and linear discriminant analysis (LDA) to learn an embedding space and was the first to reach a meaningful accuracy on a very large corpus (one million tracks). 

The impressive results of the data-driven learning approaches in other fields also fostered the use of representation learning in VI. 
A common approach became to train a convolutional neural network (ConvNet) to extract a compact representation (the embedding) out of a low- or mid-level spectral representation of the audio. 
For instance, Xu et al.~\cite{xu2018key} proposed training a ConvNet mapping PCP descriptors of each version of a composition to the same class. More generic spectral representations have also been investigated, such as the constant-Q transform (CQT)~\cite{yu2019temporal}. 
Different variants of ConvNets were proposed to address the tempo invariance problem, including temporal pyramid pooling~\cite{yu2019temporal} or standard max-pooling that have proven their efficiency in the image domain for dealing with different image scales (see Section~\ref{sec:tempotime}). These methods yielded promising accuracy and lookup times on a large corpus of tens of thousands of tracks.

\minisec{The rise of metric learning} 
Metric learning is a subset of representation learning that aims to learn a compact data representation fitting a given similarity estimation function (e.g.,~Euclidean distance). 
The underlying motivation is that the learned representations and similarity estimation functions could yield better performances for high-dimensional data, compared to their ad-hoc counterparts.
Following this idea, Doras et al.~\cite{doras2019cover} proposed learning an embedding of dominant melody, while Yesiler et al.~\cite{yesiler2019accurate} proposed an approach based on a learned PCP feature that approximates estimated chords. 
In both cases, the principle was to learn to project tracks into compact embedding vectors so that the pairwise distance (e.g.,~Euclidean or cosine distance) is smaller for versions than for non-versions.
This was typically achieved using an objective function such as a triplet loss, yielding promising results on datasets of up to fifty thousand tracks. In a similar vein, Zalkow and Müller~\cite{zalkow2020} proposed learning an embedding of short audio shingles and demonstrated the efficiency of this approach for Western classical music.

Recently, feature learning--based approaches have yielded the current state-of-the-art performance. On the one hand, a musically informed approach combining various complementary musical features, such as melody and harmony, yielded a competitive accuracy and lookup times~\cite{doras2020combining}. On the other hand, a very deep architecture applied to a generic spectral representation proved to be expressive enough to yield a similar accuracy without any prior musical knowledge~\cite{du2020bytecover}. 

\begin{figure}[tb]
\centering
\includegraphics[width=\columnwidth]{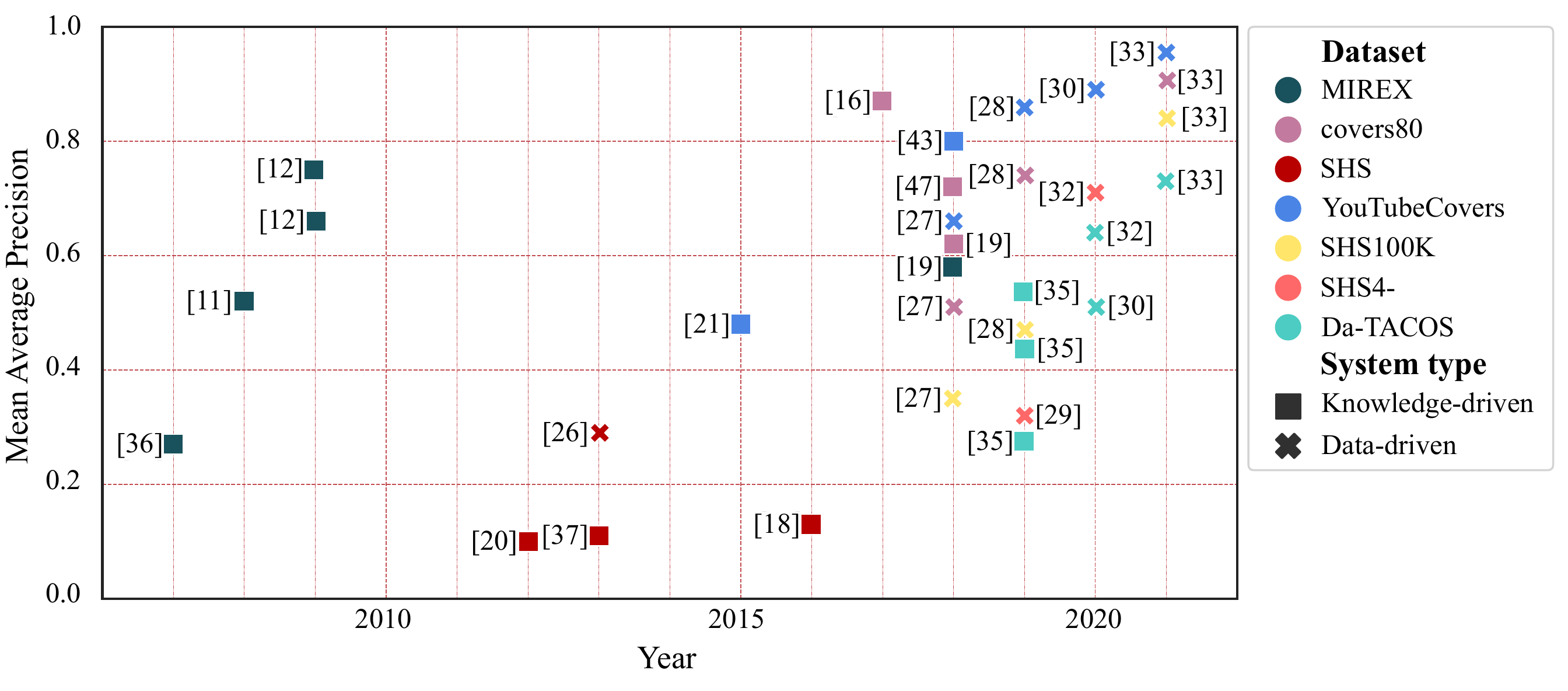}
\caption{Performance (as measured by mean average precision) of different VI systems evaluated on several datasets throughout the years.} 
\label{fig:map_summary}
\end{figure}


Throughout the last 20 years, VI systems have evolved to improve their accuracy on increasingly large corpora. This trend can be seen in Fig.~\ref{fig:map_summary}, which summarizes the mean average precision (MAP) scores that such systems obtained on different VI datasets over the years (see Section~\ref{sec:datasets} for details about these datasets and Section~\ref{sec:metrics} for details about MAP). While high performance scores could only be obtained on smaller datasets with hundreds of tracks in the early years, recent VI systems are nowadays able to reach similar performances on datasets with thousands of tracks.

\section{Building blocks of VI systems}\label{sec:blocks}
\begin{figure}[tb]
\centering
\includegraphics[width=\columnwidth]{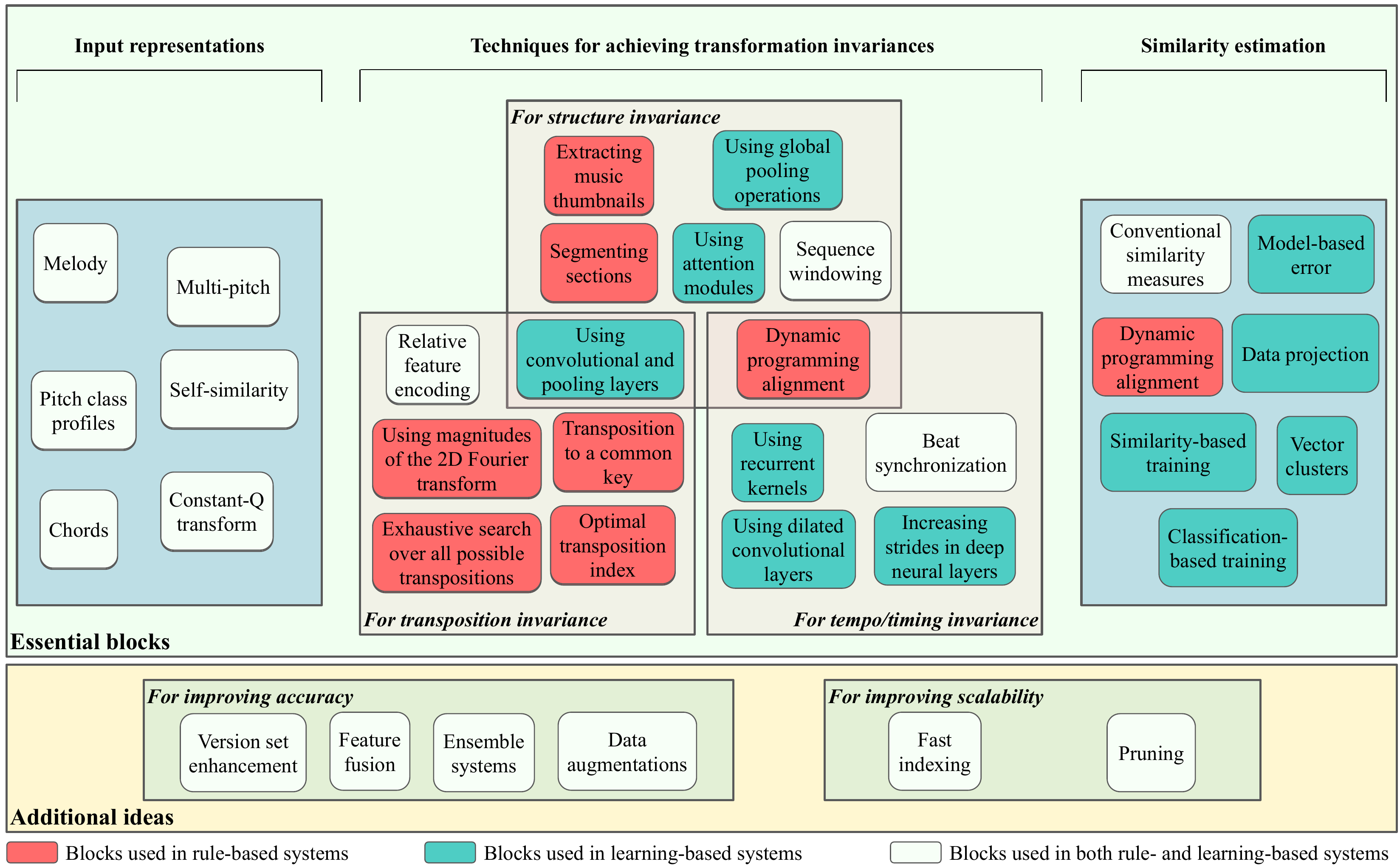}
\caption{Illustration of building blocks of VI systems detailed in Sections~\ref{sec:blocks}~and~\ref{sec:beyond_blocks}.} 
\label{fig:bblocks}
\end{figure}

Following the historical perspective presented in Section~\ref{sec:survey}, we now present a deeper dive into VI systems by dissecting them into their building blocks. Following the literature, we consider five main components that are for (1) feature extraction, (2) transposition, (3) tempo/timing and (4) structure invariance, and (5) similarity estimation (see Fig.~\ref{fig:bblocks}). While each of these blocks addresses a key challenge in the VI workflow and is proven to improve system accuracy, there is no requirement that VI systems incorporate all of them.
In fact, some of the techniques presented below may address multiple challenges at once.

\subsection{Feature Extraction}
\label{sec:input_repr}
Extracting useful information from high-dimensional audio signals is the first step in VI systems. Considering the nature of the problem, the representations that are rich in relevant characteristics (e.g.,~harmonic or melodic) and ignore the commonly varied ones (e.g.,~timbre, harmonization, or noise) are favored.

\begin{figure}[tb]
\centering
\includegraphics[width=\columnwidth]{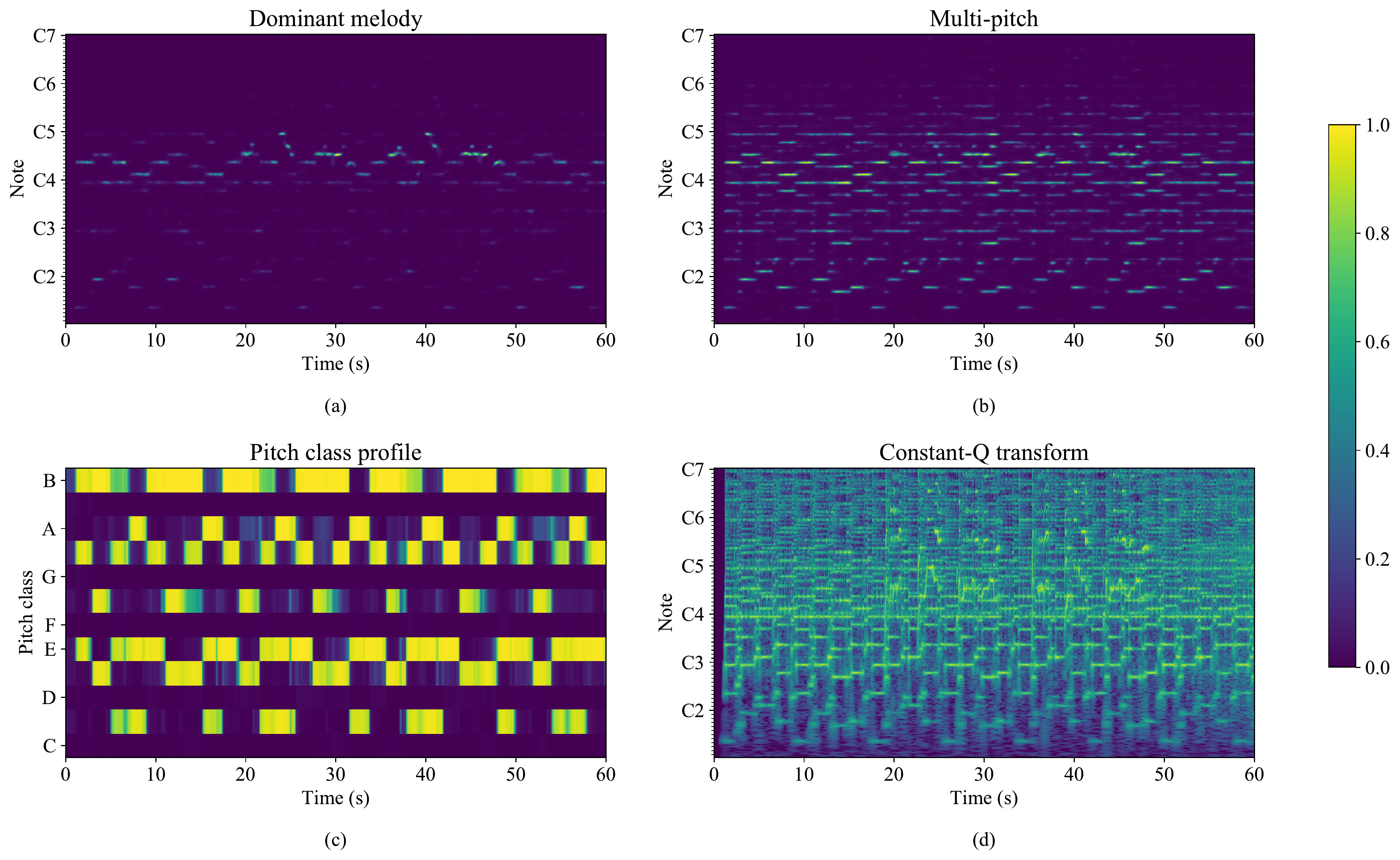}
\caption{Common input features for VI systems, extracted for the song ``Don't Stop Believin''' by Journey (included in the supplementary website). The y-axes represent musical notes (in subfigures a, b, and d) and pitch classes (in subfigure c), the x-axes represent time, and the color scale indicates the energy/intensity of such notes/pitch classes on a given time frame.}
\label{fig:spm_inputs}
\end{figure}

\minisec{Melody} Humans are very good at identifying a known song when the isolated melody is played. 
Following this intuition, melody-based representations are a natural choice for VI systems and have been explored in the literature since the early days~\cite{marolt2008mid, salamon2012melodybass, doras2019cover}. Early melody estimation systems were based on sub-harmonic summation while the recent systems are typically implemented as ConvNets.
Mainly, two types of melody representations are considered: the dominant melody, which represents a single pitch trajectory that is generated by the most dominant instrument (i.e., singing voice or a solo instrument), and the bass melody, which encodes low-frequency bass information that may be relevant for VI. An example of a dominant melody representation can be seen in Fig.~\ref{fig:spm_inputs}~(a). 
Melody representations are usually high-resolution features, both in time and frequency, in order to model the subtle variations generated by continuous pitch instruments like violin or singing voice. 
When used for VI, they are downsampled along both axes, as such levels of granularity increase computational complexity and, furthermore, incorporate detail that is detrimental to detecting variations of the same underlying musical piece.

\minisec{Multi-pitch} Another type of high-resolution feature is a multi-pitch representation, which captures information about the pitch trajectories for each source in a track, covering both melodic and harmonic content (an example can be seen in Fig.~\ref{fig:spm_inputs}~(b)).
Similar to dominant melody, multi-pitch representations are also typically downsampled along both axes and have been shown to be a useful feature for VI~\cite{doras2020combining}.

\minisec{Pitch class profile} Perhaps the most exploited musical characteristic in VI systems is the harmonic content~\cite{ellis2007identifyingcover, serra2008chroma, serra2009cross, bertin2012large, humphrey2013data, chen2018, yesiler2019accurate}. PCP has been the primary representation used to analyze the harmonic content in musical audio recordings for a long time. They are derived frame-wise by collapsing the energies within a certain frequency range (commonly 50 to 5,000\,Hz) into an octave-independent and usually 12-bin histogram that represents the relative intensities of the 12 semitones found in the Western musical tradition (see Fig.~\ref{fig:spm_inputs}~(c) for an example). 
An important variant of PCP representations is the HPCP~\cite{gomez2006automatic, serra2009cross}, which has been, and still is, used in VI extensively. 
It produces a more robust summary of the tonal content than plain PCP by incorporating additional steps such as harmonic weighting and spectral whitening. Along with HPCP, another PCP variant that has been used by many systems is chroma energy distribution normalized statistics, or CENS~\cite{muller2005audio}. 
It is obtained by incorporating quantization and smoothing operations that alleviate the issues with sensitive characteristics of local PCP distributions, namely articulation variations and local tempo deviations. 
However, the search for more robust PCP-like features is still ongoing.
A recent trend is to train neural networks to estimate ``deep'' PCP features from audio. 
For example, cremaPCP is a learned variant of PCP that estimates pitch-class information needed to predict chord sequences, and it shows performance improvements over PCP and HPCP features in the VI context~\cite{yesiler2019tacos, doras2020combining}.

\minisec{Chords} Another idea for exploiting the harmonic content is to extract chord progressions from the audio signals~\cite{bello2007audio, khadkevich2013large}. 
They can be seen as an abstraction over PCP features to obtain a more robust summary of the harmonic content, and the fact that they can be represented as discrete codes (i.e.,~chord symbols) makes them highly useful for reducing computational complexity for similarity estimations and disk usage for data storage. 
Although the motivation for using chord progressions can be easily justified, issues in chord estimation algorithms make them less appealing for VI. For example, most research in automatic chord estimation uses a rather small target vocabulary (24 chords), which is insufficient to correctly transcribe tracks from certain genres (e.g.,~jazz and blues) and may lead to inaccurate input representations. 

\minisec{Self-similarity} An interesting way to make features more robust across musical versions is to consider their evolution~\cite{tralie2015cover, tralie2017cover}.
A common way to model musical structure is via self-similarity matrices, which represent the distances between the feature vectors of each time frame and every other frame.
Self-similarity matrices are attractive input representations for VI systems because they are invariant to several musical characteristics including timbre, transposition, and noise, as they only encode relative differences between features
from different time frames, discarding their global offset.

\minisec{Constant-Q transform} For many years, the input representations for VI systems were either
mid- or high-level audio descriptors, mainly due to the fact that low-level representations contained too many redundant and noisy signals for developing an accurate system. 
However, as deep learning methods become more popular, VI researchers have begun experimenting with low-level descriptors like CQT~\cite{yu2019temporal, jiang2020}.
The CQT is a spectral representation of an audio signal, which is obtained by using a set of frequency filters with a constant-Q factor (see Fig.~\ref{fig:spm_inputs}~(d) for an example).
The representation is quite convenient for considering pitch transpositions, as the filters are logarithmically scaled and match the pitches on the Western musical scale, which is an important advantage of CQT over other spectral representations like plain short-time Fourier transform. 
Considering that many deep melody- or harmony-based representations are extracted from the CQT, it has become a natural choice for deep learning--based VI systems that follow a data-driven, feature-learning paradigm.

\subsection{Transposition Invariance}
\label{sec:transposition_invariance}
Transposing a song to a different key is a common practice in musical performances and is among the most common variations in musical versions. Thus, it is desirable for VI systems to be completely invariant to transpositions.
When not adequately accounted for, transpositions can drastically lower a VI system's performance.

\minisec{Transposition to a common key} A straightforward idea to deal with transpositions is to estimate the key of each track and transpose them to a common key (typically C~major or A~minor)~\cite{gomez2006song}. However, the accuracy of the key estimation algorithm is critical to the success of this approach, as the errors propagate to the overall system.

\minisec{Relative feature encoding} When using dominant melody or chord representations as input, instead of using the exact frequencies or chord symbols, the same information can be represented as intervals, starting from a given offset (e.g.,~the first note, the mean frequency, or the first chord of a track)~\cite{sailer2006finding, doras2019cover}.
With this, the representations are disentangled from their key offset, which results in them being transposition-invariant.

\minisec{Exhaustive search over all possible transpositions} Another approach to this problem is to obtain similarity estimations between a track and all possible transpositions of the other track~\cite{ellis2007identifyingcover, khadkevich2013large}. Especially when using PCP representations as input, their octave-independent characteristic limits the search space to 12 transpositions in total. 
This approach has the advantage of being more robust than approaches based on direct key estimation but requires a higher computation complexity in the similarity estimation step.

\minisec{Optimal transposition index} A more computationally efficient approach for considering all possible transpositions is to estimate the OTI for a pair of tracks, which indicates the 
pitch transposition interval 
needed to make the tracks at hand the most similar~\cite{serra2008chroma, serra2009cross}.
In practice, the similarity between global feature vectors of the first track and the transposed versions of the second track is estimated to find the index that results in the highest similarity.
Previous works have shown that OTI is very successful with PCP features.

\minisec{Using magnitudes of the 2D Fourier transform} The magnitude and phase components of the Fourier transform represent the energy of each sinusoidal frequency and its rotational offset, respectively. Therefore, discarding the latter provides shift-invariance with respect to the axes on which the Fourier transform is applied. 
In VI, applying a 2D Fourier transform on short patches of PCP features and discarding the phase is a practice that is used for obtaining representations invariant to pitch transpositions~\cite{bertin2012large, humphrey2013data}.

\minisec{Using convolutional and pooling layers} A typical approach for achieving translation invariance in machine learning is to use convolutional and pooling layers. 
Convolutional layers aim to capture local information with kernels that traverse the entire input, which in the case of VI are the 2D input representations presented in Section~\ref{sec:input_repr}. The output produced by convolutional layers can be aggregated using pooling layers so that the results are invariant to the exact location of a pattern of interest.
The key point for translation invariance using 
this approach
is to have kernel sizes much smaller than the input representations, so that the kernels can model similar patterns even when they are present at different locations of the input~\cite{doras2019cover}.
In order to take advantage of the octave-independent characteristic of PCP representations, convolution kernels can be combined with a simple pre-processing step and max-pooling operation~\cite{xu2018key, yesiler2019accurate}.
First, PCP features are copied and concatenated in the pitch class dimension, so that the resulting representation contains all possible transpositions for each track.
Next, a series of convolution operations of size 12 in the pitch class dimension are applied, generating activations for each transposition.
Finally, a max-pooling layer of size (12 $\times$ 1) selects the largest value across transpositions, which can be considered as the most useful transposition, for each activation.
Although proven to be useful for PCP representations, applying this technique for other input representations is not straightforward and has not been tested in the literature.

\subsection{Tempo and Timing Invariance}
\label{sec:tempotime}
Other common types of transformations in musical performances are tempo and timing changes.
Tempo differences can occur across entire tracks by changing the global speed, or within certain segments with changes in local contexts.
Timing differences often occur on the note level and consist of sustaining, repeating, shortening, or removing notes, mainly for conveying artistic expressions.

\minisec{Beat synchronization} Tempo differences result from changes in the duration of bars; however, the number of bars is not affected by such changes. 
Therefore, by using beat-synchronous features, the temporal content of the input can be represented in units of beats, rather than units of seconds (or frames)~\cite{ellis2007identifyingcover, bertin2012large}.
To compute beat-synchronous features, a beat estimation step is performed for each track, and the feature content that falls into each beat interval is aggregated to obtain one feature vector per beat. 
While the efficiency of this approach highly depends on the beat estimation algorithm, empirical results have proven this technique to be helpful for handling variations in tempo.

\minisec{Dynamic programming alignment} Perhaps the most standard way of dealing with tempo and timing modifications is to perform alignment using DP algorithms. 
Such algorithms aim to find the optimal alignment between a pair of time series given certain constraints on the solution space.
In the VI context, global alignment methods like dynamic time warping (DTW)~\cite{gomez2006song, serra2008chroma} and local alignment methods like the Smith-Waterman algorithm (SWA)~\cite{serra2009cross, chen2018} have been proven useful for achieving tempo and timing invariance.
Although resulting in higher system performances compared to other strategies against tempo and timing changes, DP methods require higher (typically quadratic) computation costs.

\minisec{Increasing strides in deep neural layers} The striding operation in neural networks is a common method for downsampling a given input, and it may be useful for being robust to timing variations~\cite{doras2019cover}. 
Although prone to aliasing, this method of downsampling applied on abstract representations found in deeper layers of neural networks is commonly seen in deep learning models, as well as some VI systems.

\minisec{Using recurrent kernels} In the machine learning community, recurrent kernels are a popular choice for dealing with sequential data. 
They often incorporate a notion of ``memory'' that facilitates differentiating between past and future events, which gives them the capacity to preserve non-stationary characteristics.
The classic formulation of recurrent kernels is prone to a number of issues while training, including the inability to consider long-term dependencies and not being suitable for parallel processing. 
Considering that the VI systems typically process full-length songs rather than short fragments, the issues around recurrent kernels make them less appealing for VI research; thus, they have been under-explored in the literature~\cite{ye2019}.

\minisec{Using dilated convolutional layers} The receptive field of convolution kernels can be increased by separating kernel elements from each other by a dilation factor while keeping the number of parameters constant. 
Using multiple kernels with different dilation rates can help efficiently 
process
a given input for various tempi using convolutional kernels~\cite{jiang2020}.

\subsection{Structure Invariance}
\label{sec:struct}
Another common source of variance across versions is the musical structure.
This includes removing, repeating, or changing the order of the existing sections, or introducing new ones.

\minisec{Segmenting sections} An intuitive solution for dealing with structural changes is to first perform segmentation in order to identify sections and later estimate the segment-wise similarities between two tracks~\cite{gomez2006automatic, cai2016}. 
Although this is an ideal solution for achieving structure invariance, segmentation algorithms are currently error-prone, and comparing wrongly segmented sections may drastically reduce the system performance.



\minisec{Extracting music thumbnails} To avoid the computational complexity of using all segments and performing segment-wise similarity estimation, some VI systems extract single or multiple ``thumbnails,'' or short representative clips, for each track and use them as a representation of it~\cite{silva2018summarizing}. Such thumbnails can be selected using various criteria, the most common being selecting the most repeated subsequences. This technique assumes (1)~that the most repeated section will correspond to the most informative or characteristic one, and (2)~that all versions of a particular song include that section due to its importance. 
While these assumptions hold true for many versions of many popular songs, extreme stylistic changes may present difficulties for this method.

\minisec{Sequence windowing} Following the idea of comparing subsequences rather than entire tracks, this technique avoids a segmentation or thumbnailing step by dividing a representation into short, overlapping segments of a fixed size (also called shingles) using a predetermined hop length between offsets of consecutive windows~\cite{muller2005audio, casey2006song, bertin2012large}. 
After obtaining multiple shingles for each input, such shingles can be aggregated by computing their mean or median~\cite{bertin2012large}, the distances obtained between an item and multiple shingles can be aggregated~\cite{zalkow2020}, or each shingle can be used individually for fragment-level retrieval.

\minisec{Local alignment} Apart from being helpful for tempo and timing invariance, alignment algorithms can also be useful for achieving structure invariance~\cite{serra2009cross, chen2018}. The key consideration for this is to avoid global alignment algorithms (e.g.,~DTW) since they struggle in cases with structural changes. Local alignment algorithms (e.g.,~SWA), on the other hand, are ideal candidates for achieving structure invariance due to their goal of finding alignments among subsequences. 

\minisec{Using convolutional and pooling layers} As introduced in Section~\ref{sec:transposition_invariance}, convolutional and pooling layers can be combined to achieve translation invariance. Such a property can be useful for achieving structure invariance as it makes the VI systems less sensitive to the exact locations of patterns (e.g.,~when the ordering of sections are changed). Moreover, to handle cases where certain sections of a track are repeated, using max-pooling for downsampling can be useful.

\minisec{Using global pooling operations} It is common practice to use local pooling operations for downsampling purposes.
Global pooling, however, takes the downsampling aspect a step further to aggregate information from an entire dimension to output a single value~\cite{xu2018key, yu2019temporal, doras2019cover}. 
The main purpose of this is to be invariant to track length and to obtain a fixed number of features per track.
Choosing the pooling operation is a crucial aspect of this technique and can be done in an intuitive way. 
The average-pooling operation considers all the temporal frames as equally important. 
On the one hand, this may hurt system performance when versions include new or missing segments, but, on the other hand, the most repeated sections will contribute to the results more than the others. 
The max-pooling operation chooses only the time frames with the highest value for each channel, and assuming that the frames with the highest value are the most informative ones (per channel), this operation is better suited against changes in structure. However, during the backpropagation phase of training, the gradients will flow only from the selected time frames for each channel, which may introduce some instability during earlier training steps due to weights being randomly initialized and updated only through such selected time frames.

\minisec{Using attention modules} Although pooling operations have been the most popular choice for structure invariance in deep learning--based VI, they consider each frame independently and ignore their relationships. 
Attention modules address this issue by inferring links between frames so that the models can select which frames to highlight or ignore based on the information contained in each frame. 
A popular attention technique in the machine learning community is self-attention, popularized by the Transformer architecture. 
Self-attention passes information about each frame to all the others and modifies the features in each time step using this information.
Ideally, only the most informative frames (e.g.,~the most repeated ones) are highlighted, and the structural changes are overlooked. 
Although widely used in the applied machine learning literature,
the self-attention idea has been under-explored in VI~\cite{jiang2020}. 
Multi-channel attention is another technique that can be considered as a link between attention and global pooling techniques. 
The goal is to learn a weighted average of time frames for each feature (or channels of convolutional layers) independently~\cite{yesiler2019accurate}.
Compared to self-attention, there are two main differences: (1) different sets of kernels are trained to learn the attention weights, which are later applied to the input of the module for performing a weighted average, and (2) using convolution kernels for computing the weights provides attention only within a local context.

\subsection{Similarity Estimation}
\label{sec:similarity_estimation}
The main goal of VI systems is to estimate similarities between pairs of tracks in a way that versions of a musical composition return higher similarity scores than non-versions. 
The techniques used for achieving invariances are crucial for this purpose. 
However, the similarity estimation algorithm must also be chosen carefully to succeed. 
Based on the literature, we consider two main types of similarity estimation: knowledge- and data-driven approaches.

\subsubsection{Knowledge-driven Approaches} Knowledge-driven approaches use heuristic-based algorithms that are often selected based on domain knowledge.
The characteristics of the invariant representations obtained from the previous steps (e.g.,~whether representations lie in Euclidean space) play an important role in the decision of which algorithm to use for this final step.

\minisec{Conventional similarity measures} When the invariant representations obtained from previous blocks are suitable, similarity measures such as cross-correlation~\cite{ellis2007identifyingcover}, the Euclidean distance~\cite{marolt2008mid}, or the dot product~\cite{muller2005audio} are simple, yet effective choices for similarity estimation. 

\minisec{Dynamic programming alignment} As described in Sections~\ref{sec:tempotime} and~\ref{sec:struct}, DP techniques are often used for tempo, timing, and structure invariance, and they eliminate the need for adopting further similarity estimation steps as they 
provide a measure of it. 
In the cases of alignment algorithms like DTW, the cost of the optimal solution can be used as a distance measure~\cite{foote2000arthur, gomez2006song, gomez2006automatic}. 
In the case of local alignment algorithms like SWA, the length of the longest-aligned subsequence is typically considered as a measure of similarity~\cite{serra2009cross, chen2018, tralie2017cover}.

\subsubsection{Data-driven Approaches} Data-driven approaches aim to learn a function that transforms the data 
to facilitate similarity estimation through conventional measures (e.g.,~Euclidean distance).
Such functions are learned using training data, in a supervised or unsupervised fashion.
In supervised cases, the semantic relationships (i.e.,~version or non-version) between pairs of items are used for obtaining effective similarity functions.
The learned functions depend on the inductive biases of the models and the training process.

\minisec{Data projection}
Along with a few classification approaches, early works for data-driven VI considered 
data projection
algorithms like PCA and LDA~\cite{bertin2012large, humphrey2013data}. 
They are used for transforming invariant representations obtained in the previous blocks into more compact embedding vectors. 
Such systems can be considered as hybrid approaches that connect knowledge- and data-driven similarity estimation, as they incorporate rule-based algorithms for their initial steps.

\minisec{Vector clusters} Another common approach to 
derive
input representations was based on vector clustering, such as $k$-means. A learned dictionary of cluster centroids can be used to efficiently encode data with a small number of components. This approach has been used to encode input representations into chord series~\cite{bello2005robust}, hashcodes~\cite{kurth2008efficient}, or embedding vectors~\cite{humphrey2013data}.

\minisec{Model-based error} Another early data-driven approach to VI was to study model-based errors~\cite{serra2012, foster2015}. 
In this approach, a simple parametric model that describes the temporal evolution of the feature sequence is fit to the data.
This modeling can be performed on the basis of a single musical piece or from multiple pieces that form a version clique.
After that, the model is used to predict future samples of a new feature sequence coming from a candidate piece (i.e.,~fragments of the sequence are used as input to the model, and this outputs the most reasonable continuation).
If the model produces a small error, one concludes the candidate piece is a version of the piece that was used to train the model.

\minisec{Classification-based training} Classification-based training approaches are perhaps the most popular in supervised learning. 
In VI, three main formulations exist. 
Firstly, distance/similarity scores can be obtained from multiple systems and used to train a classifier to make a final decision (see Section~\ref{sec:ensemble_systems}). 
Secondly, a cross-similarity matrix of two tracks can be computed, and a convolutional network used to determine whether the inputs are versions of each other or not (i.e.,~binary classification)~\cite{lee2018}. 
Although computing cross-similarity matrices introduces a computational load for the similarity estimation step, convolutional networks can replace the quadratic-complexity alignment algorithms like SWA, which results in a considerable improvement in terms of overall computational requirements. 
Thirdly, the training process can be formulated by considering each clique as a separate class (i.e.,~multi-class classification)~\cite{xu2018key, yu2019temporal, jiang2020}. 
One important consideration is that, during inference, it is likely that the system will encounter tracks from cliques that are not in the training data, and using a pure classification strategy, it would not be possible to correctly identify those cases. 
To handle this, a typical solution is to consider the output of the penultimate layer of the model as the embedding for each track.
Training then aims to make the embeddings from different classes linearly separable, which is a way of constructing a similarity function.

\minisec{Similarity-based training} In recent years, the most popular training formulation in VI is similarity-based, using metric learning approaches~\cite{doras2019cover, yesiler2019accurate, jiang2020}. 
The supervision signals in this context only require the information about whether two items are similar (as in, belong to the same clique) or not. 
During training, instead of predicting the classes of each item, these systems focus on manipulating the distances of items directly by pulling together similar items and pushing apart the dissimilar ones. 
The most popular training objectives for this approach are contrastive and triplet losses.

\section{Beyond building blocks: improving accuracy and scalability}\label{sec:beyond_blocks}
This section introduces a set of ideas that can be incorporated in VI systems regardless of their building blocks, mainly for improving their accuracy or scalability.
We group these ideas into six main categories: version set enhancement, feature fusion, ensemble systems, pruning, fast indexing, and data augmentation.

\subsection{Version Set Enhancement} 
\label{sec:clique_enhancement}
Versions of the same composition can be viewed as items of the same set, or clique. Using this intuition, community detection algorithms have been studied to refine the obtained distances between queries and items in a corpus. 
Specifically, one can construct a fully connected graph based on similarities obtained with any system, eliminate certain edges based on a threshold of some quantity (e.g.,~a distance threshold), and, assuming transitive relations, complete the missing links in this graph~\cite{serra2009unsupervised}. 
This process is highly efficient and can lead to better retrieval performance by finding undetected versions and cleaning up the noisy results. 
As a side benefit, it is possible to use measures of centrality on these completed graphs to estimate the original performance from which subsequent versions arose~\cite{serra2009unsupervised}.

\subsection{Feature Fusion} 
\label{sec:feature_fusion}
Considering the complexity of the VI task, no single feature has been able to capture all possible transformations that exist across versions. 
For instance, while the majority of VI systems work by matching sequences of pitch-based features, this leaves a blind spot for certain genres where the notes do not carry the dominant musical expression, such as '80s hip-hop and drum solos~\cite{tralie2017cover}. 
At the same time, features that ignore notes are missing crucial information that helps much of the time~\cite{tralie2015cover}. 
Hence, intuitive solutions for such issues have been proposed to find ways of combining information from various musical dimensions.

\minisec{Early fusion}
Music, in general, has repeated structures that can be represented as a graph, where each node represents a small snippet of audio and edges exist with high weights between snippets that are similar, according to some chosen features.
Different features can lead to different noisy observations of an ideal structure graph and, in most cases, no single feature reliably picks up on all aspects of the 
complex musical
structure.
To address this, it is possible to use a technique known as similarity network fusion~\cite{tralie2017cover, chen2018} to reconstruct a cleaner graph from these noisy observations, particularly if they contain complementary information.
It is possible to adapt these features so that cleaner cross-similarity measures can be obtained between versions, and this can significantly improve the system accuracy over each feature alone~\cite{tralie2017cover}.

\minisec{Late fusion}
Another possibility for feature fusion is to combine information from various features at later stages of systems. 
For example, cross-similarity matrices for the same pair of tracks but obtained with different features can be aggregated using simple schemes like taking the maximum or the minimum~\cite{foucard2010multimodal}. 
In addition, embedding vectors obtained with systems that use different features can be concatenated and projected into a new space, which can then be shaped by combined characteristics of all input features~\cite{doras2020combining}.
 
\subsection{Ensemble Systems}
\label{sec:ensemble_systems}
One common strategy to boost overall accuracy is to incorporate the output from multiple pipelines. 
Such systems, known as ensemble systems, thereby leverage the joint strength of disparate workflows. 
Although the motivation behind some ensemble systems is similar to that of feature fusion (combining information from various musical dimensions), here, we describe VI systems that combine multiple systems after they return distance or similarity scores between queries and a corpus of tracks.

\minisec{Training a classifier}
A first approach for aggregating 
scores obtained from various systems is to train a shallow classifier that takes a set of scores as input and returns a binary decision (i.e.,~version/non-version). 
Depending on the characteristics of the classifier, non-linear relationships between the input scores may be explored. 
In VI, this strategy has been explored to combine systems that use different input features (e.g.,~PCP, dominant melody, and bass line)~\cite{salamon2012melodybass}, and different similarity estimation steps (e.g.,~local alignment and cross-correlation)~\cite{ravuri2010cover}.

\minisec{Similarity normalization and aggregation}
In cases where the distance scores obtained from various systems are well-calibrated, aggregation of such distances can be trivial with simple schemes like taking the mean, the maximum, or the minimum.
However, when there is a mismatch regarding the scale of such scores, additional operations like simple normalizations are needed to alleviate the issue~\cite{degani2013heuristic}.

Another approach to handle scores of different scales is to consider the global ranking of all tracks in a corpus for each system since rankings are automatically invariant to the scale of the similarity scores obtained from a system.
Rank aggregation can then be used to create a 
global ranking that incorporates the individual ranks from each system. 
The Kendall tau distance~\cite{osmalsky2016enhancing} is an objective function for measuring the agreement of rankings and indicates the number of pairs of ranked tracks that have a reversed order.
A Kemeny optimal global ranking~\cite{osmalsky2016enhancing} is a ranking that minimizes the Kendall tau distance to each individual system's rank. 
Unfortunately, finding such a ranking is NP-hard. 
However, using an initial guess based on heuristics such as mean and median rank aggregation, followed by local Kemenization, in which greedy swaps are performed until the Kendall tau distance is minimized, can lead to superior performance over individual systems in practice~\cite{osmalsky2016enhancing}.

\minisec{Late similarity network fusion}
In addition to promoting cliques 
from a single system, it is also possible to fuse graphs from multiple similarity networks. 
Similarity network fusion can again be used 
to enhance cliques, but 
the algorithm operates at the track level instead of the time frame level (as done in feature fusion). 
Due to normalizations based on local neighborhoods within each network, this technique can fuse similarity measures from any set of systems, and it has been shown in practice to improve accuracy when fusing 
PCP-based systems that use different alignment schemes~\cite{chen2018}, as well as between systems built on timbral and PCP-based features~\cite{tralie2017cover}.

\subsection{Pruning} 
\label{sec:pruning}
As many information processing systems suffer from the accuracy--scalability trade-off, a plausible solution is to design multi-step systems where the scalability and the accuracy of multiple systems may complement one another.
For this, fast algorithms can prune the corpus to allow slow but better-performing systems to operate on a reduced set of data for improving the computation times.

\minisec{Scalable VI systems}
Lately, deep learning--based VI systems have made substantial contributions for bridging the accuracy--scalability gap, but before them, scalable VI systems were not sufficient for obtaining confident results. 
However, considering their far-from-random performances, such early systems became a natural choice to be used as the first step of pruning-based, multi-step systems~\cite{cai2016}. The general tendency was to use systems that encode tracks into compact embedding vectors as the first step, mainly to take advantage of the fast lookup times. 
Afterward, local alignment--based systems were used on the pruned set of tracks to obtain the final results. 
Pruning systems nonetheless need to have good recall (at the expense of good precision, if necessary).

\minisec{Weak rejectors}
There is a multitude of features that are similar for versions of the same track, but which are not strong enough indicators to confidently label them as versions of one another. 
Still, having a large collection of such ``weak rejectors'' can be used to narrow down candidates, leading to improved scalability.
Among such features are bag of words of PCP features, duration and tempo of a recording~\cite{osmalsky2016enhancing}, and structure-related descriptors~\cite{yesiler2019tacos}. 
Although not applicable in the audio-based VI literature,
textual bag of words can also be applied to title and lyrics information, if they are available~\cite{correya2018large}.

\subsection{Fast Indexing} 
\label{sec:fast_indexing}
Like every other information retrieval system, VI systems store and index tracks for future lookup and comparison. 
This perspective motivated other strategies to devise new music representations, inspired by text-based content indexing. 
Different kinds of efficient text indexing algorithms were then adapted to music retrieval: for instance, and among others, inverted file indexing and LSH. 

\minisec{Inverted file indexing} In a text-based indexing context, the idea is to establish a list of keywords (the codebook) and to use these keywords to index the documents where they appear. 
In a VI context, the codebook contains encodings of audio shingles, which are used as indexes to the full audio tracks. 
For instance, a $k$-means approach was proposed to learn a codebook from the set of all PCP vectors present in a corpus. 
The inverted index was built using the closest code to each PCP frame as an index to the full song~\cite{kurth2008efficient}. 
Another proposal built the codebook encoding each PCP sequence as a major/minor chord series and then using short chord subsequences as index entries~\cite{martin2012blast}.

\minisec{Locality-sensitive hashing} The basic idea of LSH indexing is to devise a hashing function that will guarantee that similar contents are encoded by the same hash with a high probability, while the probability that different contents are mapped to the same hash remains low.
There are various ways to generate hashing functions that will satisfy these properties~\cite{casey2006song}.
Several authors adapted this principle to the VI context and proposed encoding shingles of input representations with an LSH scheme (e.g.,~using dominant melody~\cite{marolt2008mid}, chord progression~\cite{khadkevich2013large}, or PCP~\cite{casey2006song}).

The recent deep learning--based systems (see Section~\ref{sec:similarity_estimation}), which also encode tracks into compact vector embeddings, have superseded these fast-indexing approaches. 
However, techniques like LSH could still be considered to further speed up the retrieval process of deep learning--based systems (see Section~\ref{sec:scalability_tradeoffs}).

\subsection{Data Augmentation} 
With the increasing interest in data-driven VI systems, domain-specific strategies for robust representation learning are becoming more important. 
For this, we now introduce data augmentation strategies that are inspired by musical characteristics that can be modified while creating versions of a composition.

\minisec{Pitch transposition}
To simulate pitch transpositions that are typically observed between versions, different strategies can be used based on the input representation. 
Firstly, regardless of the input representation, a pitch shift transformation can be applied to the audio signal before feature extraction. 
While this operation is universal and does not depend on the type of input, it can introduce some artifacts on the signal and may require more computational resources than its alternatives. 
Secondly, if the PCP features are used as input representations, a simple circular shift along the frequency axis is sufficient to simulate a pitch shift operation, thanks to their octave-independent characteristics~\cite{xu2018key, yesiler2019accurate}. 
Lastly, if melody or CQT representations are used, shifting the values by a certain number of rows along the frequency axis can be useful. 
However, unlike PCP features, these representations are not octave-independent, and the behavior of this operation at the boundary bins (the lowest and the highest) should be considered.

\minisec{Tempo}
As with pitch shift transformations, increasing or decreasing the tempo of a track can be done 
before the feature extraction step.
A common alternative to this is to apply interpolation functions (e.g.,~linear) to the 2D input representations (e.g.,~PCP, melody, or CQT)~\cite{doras2019cover}.

\minisec{Timing}
To simulate minor timing variations where some notes are sustained, repeated, shortened, or removed, similar operations can be applied to randomly selected frames from 2D input representations. 
For this, such frames can be duplicated, silenced (by replacing them with zero vectors), or simply removed~\cite{yesiler2019accurate}.

\minisec{Input patch sampling}
Similar to the idea of shingling, the input patch sampling strategy is to randomly select fixed-size patches from the input representations to use in the training process~\cite{yesiler2019accurate, yu2019temporal}. 
This can be viewed as simulating structural changes where some sections are removed from a version. Note that the sizes of the patches for this strategy (e.g.,~120--180\,s) are generally larger than the sizes used for shingling (e.g.,~30--60\,s).

\minisec{Duration}
While applying the input patch sampling transformation, the sizes of the patches can be varied to mitigate bias toward a certain representation length~\cite{yu2019temporal}.

\minisec{Noise}
Lastly, several transformations to audio signals can be applied to simulate the differences in recording conditions. Some examples are additive noise, low-pass filters, and MP3 transcoding.

\section{Datasets and Evaluation Metrics}\label{sec:data_and_eval}
This section presents an overview of the publicly available datasets and the most widely used evaluation metrics for VI. Although there exist different datasets and evaluation methods for various subproblems within VI (see Section~\ref{sec:sub-fields}), we here focus only on the most frequently used ones.

\subsection{Datasets}\label{sec:datasets}
Finding data for developing and evaluating MIR systems is a challenging issue, mainly due to the fact that musical audio is often subject to copyright. 
Historically, the impact of this issue on VI was that the researchers were limited to developing and evaluating their systems using in-house private corpora, which made unified benchmarking of systems a difficult task. 
These datasets had varying characteristics, such as the size of the corpus, the cardinality of cliques, the distribution of musical genres, and so on. 
However, with the help of online communities like SecondHandSongs.com, where editors and users annotate musical versions in terms of their connections with previous musical compositions, this issue is mostly alleviated today. 
Therefore, for the remainder of this section, we focus only on the public benchmarks and publicly available datasets that have been frequently used for VI. 
A summary of such datasets can be seen in Table~\ref{tab:vi-datasets}.

\begin{table}[tb!]
\caption{Publicly available VI datasets}
\label{tab:vi-datasets}
\begin{threeparttable}
\resizebox{\textwidth}{!}{%
\centering
\begin{tabular}{l p{0.15\linewidth} p{0.15\linewidth} p{0.15\linewidth} p{0.35\linewidth}}
\hline\hline
\textbf{Dataset} & \textbf{Training subset\tnote{1}} & \textbf{Validation subset\tnote{1}} & \textbf{Test subset\tnote{1}} & \textbf{Content}   \\
\hline

MIREX collection & - & - & 330 (30) + \newline 670 noise tracks & Proprietary collection \\
covers80~\cite{ellis2007covers80}& - & - & 160 (80) & Full audio tracks and metadata  \\

SecondHandSongs~\cite{bertin2011million} & 12,960 (4,128) & - & 5,236 (1,726) &
Pre-extracted features (a wide range including PCP, timbral features, beat, etc.) and metadata \\

%
YouTubeCovers~\cite{silva2015music} & 100 (50) & -  & 250 (50) & 
Pre-extracted features (3 PCP variants) and metadata \\

%
SHS-100K~\cite{xu2018key, yu2019temporal} & 84,340 (4,611) & 10,883 (1,842) & 10,547 (1,692) & YouTube URLs and metadata  \\
Da-TACOS~\cite{yesiler2019tacos} & 83,904 (14,499) & 14,000 (3,500) & 13,000 (1,000) +  \newline 2,000 noise tracks & 
Pre-extracted features (3 PCP variants, timbral features, and 4 rhythm features) and metadata \\
SHS5+ \& SHS4-~\cite{doras2019cover} & 62,311 (7,460) & - & 48,483 (19,445) & Pre-extracted features (CQT, 2 melody, and PCP variants) and metadata\\
\hline\hline
\end{tabular}%
}
\begin{tablenotes}
\item[1] Values outside and inside the parentheses indicate the number of tracks and unique cliques, respectively.
\end{tablenotes}
\end{threeparttable}
\end{table}

\minisec{MIREX collection} The ``audio cover song identification'' competition in MIREX stood out as the only platform for benchmarking in the early days of VI. The dataset used in this competition is private and includes 1,000~tracks from a variety of genres. 670~among them are considered as ``noise’’ tracks that do not belong to the same clique as any others. The rest of the data is organized into 30~cliques with 11~versions each. While the query set consists of only those with multiple versions (330~tracks), the corpus includes the entire collection of 1,000~tracks. The inclusion of noise tracks that are not queried is done mainly to imitate the distribution of industrial corpora, and it influenced some of the forthcoming publicly available VI datasets.

\minisec{covers80} Apart from the MIREX collection, the first dataset curated for VI research is covers80~\cite{ellis2007covers80}, and it was released publicly as opposed to the former. It includes full-length audio files for 160~tracks divided into 80~cliques with 2~versions each and no noise tracks. The majority of this data is taken from the uspop2002 dataset\footnote{\url{https://labrosa.ee.columbia.edu/projects/musicsim/uspop2002.html}}, and the rest was taken from a few commercial ``cover albums.'' The major advantage of this dataset is that it includes audio files for all tracks, which enables researchers to develop and evaluate systems that use novel input representations. However, the limited size is an important drawback as the reported results may not be statistically significant for a true comparison of systems.

\minisec{SecondHandSongs} The next publicly available dataset for VI was the SecondHandSongs (SHS) dataset~\cite{bertin2011million}. It is a subset of the Million Song Dataset~\cite{bertin2011million}\footnote{\url{http://millionsongdataset.com/}}, and the version annotations are obtained using the SecondHandSongs.com API\footnote{\url{https://secondhandsongs.com/page/API}}. It includes a training set with 12,960~tracks split into 4,128~cliques and a test set with 5,236~tracks split into 726~cliques, without any noise tracks. With the release of SHS, VI research
entered into a new era, which led to development of scalable systems that can leverage and be evaluated on large datasets. However, due to legal issues, this dataset includes only pre-extracted features that were obtained using the EchoNest API\footnote{\url{https://en.wikipedia.org/wiki/The_Echo_Nest}}. Therefore, the VI systems developed and evaluated using this dataset have a strict limitation in the input representations they can use, which may reduce accuracy and hinder system deployment in the real world due to their proprietary nature.

\minisec{YouTubeCovers} Following the idea of sharing pre-extracted features, the YouTubeCovers dataset was released with a larger set of harmonic features compared to SHS~\cite{silva2015music}. It includes a total of 350~tracks from 50~cliques and is further split into a training subset with 100~tracks (2~per clique) and a test subset with 250~tracks (5~per clique) with no noise tracks. However, having the same cliques in both training and test subsets may result in biased evaluations. Moreover, like covers80, the rather small evaluation set may lead to statistically insignificant results. 
Although there are still research papers using this dataset, the URL shared in the original publication for obtaining the dataset is no longer maintained. 

\minisec{SHS-100K} With deep learning--based systems getting more prominent, a need for larger datasets has emerged. Addressing this need, SHS-100K includes a total of 108,869~tracks split into 9,202~cliques (no noise tracks), which was a considerable increase compared to the largest dataset until then~\cite{xu2018key}. The version annotations are collected from SecondHandSongs.com, and the dataset includes YouTube links for the tracks, rather than any pre-extracted features. It was initially divided into training, validation, and test subsets with 101,968~tracks (8,177~cliques), 3,918~tracks (909~cliques), and 2,983~tracks (116~cliques), respectively. However, in a later publication, the dataset was split in a different way mainly to have a larger test set, having 84,340~tracks (4,611~cliques), 10,883~tracks (1,842~cliques), and 10,547~tracks (1,692~cliques) for training, validation, and test, respectively~\cite{yu2019temporal}.

\minisec{Da-TACOS} Another dataset that addressed the need for larger corpora is Da-TACOS~\cite{yesiler2019tacos}. It includes a benchmark set with 15,000~tracks that split into 1,000~cliques with 13~versions each and 2,000~noise tracks. Like many others, the version annotations are obtained using the API of SecondHandSongs.com. To enable researchers to experiment with not only harmonic but also rhythmic and timbral characteristics, it includes a large set of pre-extracted features along with the metadata that is linked to the composition and performance IDs used in SecondHandSongs.com. Therefore, even though the audio files are not available, researchers can use the detailed metadata to recover the tracks themselves. Along with the benchmark set, the authors also released a framework called ``acoss'' for feature extraction and benchmarking designed for VI. It includes feature extraction functions with the hyperparameters used for preparing Da-TACOS, and open-source implementations for 7~VI systems. Furthermore, a training set for Da-TACOS was recently released, containing a training subset with 83,904~tracks (14,499~cliques), and a validation subset with 14,000~tracks (3,500~cliques). 

\minisec{SHS-5+ \& SHS-4-} The last dataset we introduce in this section is SHS-5+ \& SHS-4-~\cite{doras2019cover}. It includes a training subset, SHS-5+, with 62,311~tracks in 7,460~cliques, and all the cliques have at least 5~versions each (hence the name). The test subset, SHS-4-, includes 48,483~tracks in 19,445~cliques, with 2 to 4~versions per clique, and it is the largest benchmark set for VI to date. Neither subset includes noise tracks. The version annotations are obtained using the SecondHandSongs.com API, and the dataset includes a large set of pre-extracted features, including CQT and melodic representations. 

\subsection{Evaluation Metrics}\label{sec:metrics}
As previously introduced, VI systems aim to model 
the shared information between
versions of the same underlying composition in order to provide a similarity score. The ability of a system to correctly assess the similarities between a query track and a corpus of tracks
is usually evaluated by metrics that operate on a ranked sequence of results. Such a ranked sequence is obtained by first estimating similarity scores between a query and all the tracks in a corpus and then sorting the items in the corpus with respect to their similarities to the query. Below, we introduce the set of metrics typically used in VI (see~\cite{doras2020phd} for details). 

\minisec{Precision and recall} Being perhaps the most typical metrics in information retrieval, precision and recall provide rank-independent measures of system performance, which means that the order of the items does not affect the outcome. Precision gives the ratio of the retrieved items that are relevant\footnote{Here, we use the term ``relevant'' to denote the items in the corpus that are versions of the query.} to all the retrieved ones (in other words, the accuracy of the system's predictions). Recall, on the other hand, gives the ratio of the retrieved items that are relevant to all the relevant items in the corpus (in other words, how well the system finds the relevant items). 
In VI, they are typically computed as Precision@$K$ (P@$K$) or Recall@$K$ (R@$K$) at a cut-off rank $K$, meaning only the first $K$ results are considered. Note that although precision and recall are rank-independent (i.e.,~the order of the items does not matter), P@$K$ and R@$K$ require a cut-off rank by definition and can be considered as rank-aware (i.e.,~the items have to be placed below rank $K$).

\minisec{Mean average precision} Although precision and recall are common metrics, their rank-independent characteristics are not useful for tasks where the order of the retrieved items is important. A possible alternative for considering the ranks of the results is mean average precision (MAP). For this, average precision (AP) for all queries are computed and averaged. AP for a single query is obtained by averaging P@$K$ scores over all $K$ where a relevant item is returned, which makes AP a rank-aware metric in contrast to precision. Therefore, MAP assesses not only the number of relevant items in the results but also their ranks.

\minisec{Mean reciprocal rank} Another rank-aware metric for assessing system performances is mean reciprocal rank (MRR). It is the average of reciprocal rank scores obtained for all the queries, where reciprocal rank is the multiplicative inverse of the rank of the first relevant item. 
Therefore, this metric is more appropriate for cases where either there is only one relevant item in the corpus, or if only the position of the first relevant item is important. Note that when there is a single relevant item in the corpus, MRR is equal to MAP.

\minisec{Mean rank of the first relevant item} The last metric we introduce is the mean rank of the first relevant item (MR1). As MRR, this metric also uses only the first relevant item, and the only difference between them is the multiplicative inverse function in MRR. However, MR1 may be easier to interpret as ranks are taken directly. Note that the scale of differences between MR1 scores are the same everywhere, but such differences between MRR scores are scaled down when 
higher
ranks are considered. For example, when comparing two cases where the first relevant items have ranks 1 and 11, the 
reciprocal ranks
are 1.00 and 0.09, respectively. On the other hand, when comparing two cases where the ranks of the first relevant items are 40 and 50, the 
reciprocal ranks
are 0.025 and 0.020, respectively.

\section{Open issues and future directions}\label{sec:issues}
Although VI research has made substantial progress in the last 20 years, there are many challenges that have yet to be addressed.
Here, we outline some current open issues to provide guidance for researchers that are interested in contributing to the field.
These challenges include, but are not limited to, (1) the task definition itself, (2) evaluation methodologies, (3) trade-offs that arise when scaling VI systems, (4) accuracy gaps on certain under-represented types of content, and (5) the variety of applications and the different treatments they require.

\subsection{Task Definition}
As discussed in Section~\ref{sec:intro}, the definition of a musical version, especially a ``cover song,'' may differ in various contexts. To avoid such differences, we have so far favored a quite permissive definition in this article, labeling all the tracks that are derived from a musical composition as versions. Although this definition is convenient for introducing and discussing VI from an academic research point of view, applications that consider legal aspects of VI (e.g.,~detecting copyright infringements) may require different definitions that are more suitable for their purposes.

The definition of a version for legal applications often needs to be based upon the rightsholders (typically songwriters and recording labels) rather than the musical connections.
For example, a composer may copyright a new arrangement of a folk song that is in the public domain. According to our inclusive definition of a version, the arrangement would be a version of the original, but legally, it may be a separate entity.
These rights themselves are often not well-defined, and there are a number of famous lawsuits\footnote{\url{https://www.bbc.com/culture/article/20190605-nine-most-notorious-copyright-cases-in-music-history}} about whether or not the creators of a track must pay royalties to the rightsholders.
Publishing data, which provides a legal link between track metadata and composition metadata, often exists in text form, linking songwriters/composers, track titles, and recording/composition identifiers such as ISRCs/ISWCs\footnote{\url{https://isrc.ifpi.org/}}.

There are cases where this metadata is the only information separating two nearly identical tracks (such as an original release and a remastered release) into different legal entities.
Thus, purely audio-based VI for legal applications is not possible in many cases,
and any successful system must consider additional information such as editorial metadata to disambiguate unclear cases.

Apart from the legal perspective, current VI systems are typically built around a particular notion of which musical features are important for determining whether two tracks are versions of one another.
These notions fit well for some music traditions, such as Western pop and classical music; 
however, there are other musical traditions that break some of these assumptions.
For example, in Indian art music, certain melodic phrases or motifs are crucial for identifying the ragas (roughly speaking, modes) that tracks belong to. From a Western point of view, two tracks that include versions of the same melodic phrase would mostly mean that they originate from the same composition; however, in Indian art music, it can simply mean that they belong to the same raga.

\subsection{Evaluation Methodologies} 
Evaluation of VI systems is typically performed on well-curated datasets (see Section~\ref{sec:datasets}) using mostly rank-aware evaluation metrics (see Section~\ref{sec:metrics}).
These datasets are dominated by music with singing voice, music with well-defined versions, music from mostly pop, rock, and jazz genres, and generally do not contain duplicates. Industry-scale music corpora, however, often have quite different distributions. In such corpora, near-exact duplicates are very common,
it is difficult to apply the concept of versions for the majority of music,
genres like rap and electronic music constitute a large proportion of the data,
plus a non-negligible subset of the tracks does not contain singing voice.
This presents a challenge when extrapolating the performance of a system on an industry-scale corpus using evaluations performed on well-curated datasets. Furthermore, there are several limitations of the commonly used evaluation metrics. Firstly, they are highly sensitive to the number of relevant tracks in the corpus per query, which may not be appropriate for VI since some compositions may have hundreds of versions while some others may have only a few, or even zero. Additionally, the metrics mostly consider only the rank ordering of the tracks and not the distances between the query and the retrieved items, which makes it difficult to assess how well-separated the relevant items are from the irrelevant ones. 

Near-duplicates (i.e.,~content that is the same but distorted enough so that a standard music fingerprinting algorithm will not provide a match) present a particular challenge: 
given a query, VI systems will naturally assign a lower distance to a near-duplicate compared to other versions that may contain several changes in musical characteristics. 
This highlights both of the previously mentioned issues regarding the evaluation metrics.
For example, for P@$K$, the presence of duplicates can both increase and decrease the metric in an unpredictable way, as this changes the number of relevant tracks for a given query.
Consider this toy example: corpus A has no duplicates, and corpus B 
is an extended version of corpus A, where each item has one near-duplicate. 
Now consider a VI system that, for a given query, would return 5 relevant results out of the first 10 for corpus A (i.e.,~P@$10=0.5$).
If the relevant results are in positions 1 through 5, the equivalent query for corpus B would have 10 relevant results (i.e.,~P@$10=1.0$).
Conversely, if the relevant results are in positions 6 through 10 (for corpus A), the equivalent query for corpus B would have 0 relevant results (i.e.,~P@$10=0.0$).

Music without well-defined versions, such as ambient music and soundscapes, leave an open question: how should VI systems handle this type of content?
Practically, a VI system should not retrieve any matches when no versions are present.
However, this kind of content typically does not have 
clear melodic, harmonic, or structural characteristics.
As a result, the features VI has historically used are typically close to zero everywhere and are confidently, and incorrectly, clustered together as a tight group by VI systems.

The genre distributions of research datasets may introduce a bias toward certain input representations and musical characteristics. For example, the success of PCP representations in VI is fairly easy to explain for datasets having many tracks from the pop, rock, and jazz genres. However, the performances of state-of-the-art systems on other genres where rhythmic and timbral properties are highlighted is an under-explored issue in VI. Considering the popularity of hip-hop and electronic music genres (and their sub-genres), this is clearly an issue to be addressed in VI research to be useful in the current music ecosystem. 

Finally, VI research, except for a subfield focusing on Western classical music, has under-explored how systems behave for instrumental music, largely due to the existing datasets being dominated by music with singing voice. 
As a result, the performance of current VI systems on instrumental music is not well understood, and thus the performance on industry-scale corpora, which contain a considerable percentage of instrumental content, cannot be inferred.

\subsection{Scalability Trade-offs} 
\label{sec:scalability_tradeoffs}
Although VI has clear industrial applications in the current music ecosystem, the scope of scalability-related discussions in research papers is rather limited. It has been demonstrated that vector-based techniques provide large benefits in computation and memory requirements compared to alignment-based ones, but no systematic evaluation of vector-based techniques has been performed from a scalability perspective. The scalability considerations in such works are typically limited to the size of the embedding vectors. However, the computations that produce these vectors (e.g.,~feature extraction algorithms, or deep neural network layers) are mostly ignored. Therefore, here, we highlight two directions to cover this under-explored perspective.

Firstly, for the vector-based techniques, in particular, there is an additional accuracy--scalability trade-off that arises especially in industry-scale datasets: retrieving the $K$ nearest results for a query.
Consider a vector-based system, with vectors in $\mathbb{R}^d$ under the Euclidean norm. 
In order to retrieve the $K$ nearest results from a corpus of $N$ items in an exact manner, at least $O(N\log{N})$ operations are required, and these exact algorithms are difficult to improve due to the ``curse of dimensionality.''
Even if pruning techniques are employed to reduce $N$, when performing large numbers of $K$ nearest neighbor look-ups, the computational efficiency can greatly influence the speed and cost of deploying a VI system.
In practice, approximate nearest neighbor search algorithms\footnote{For example, \url{https://github.com/erikbern/ann-benchmarks}} are used.
These algorithms provide efficient approximations for finding the $K$ nearest neighbors for high-dimensional data over large corpora and introduce a trade-off between the time-per-query and the recall (i.e.,~the percentage of true nearest neighbors returned): the higher the recall, the slower the query.
The degree of trade-off varies by algorithm and dataset, but roughly, at 80\% recall, these algorithms provide a speedup of a factor between 100 and 1000 over an exact computation. 
For a fixed speedup, the achieved recall typically decreases as the dimensionality of the vectors increases, which highlights an additional motivation for the dimensionality reduction or data projection methods described in Section~\ref{sec:similarity_estimation}.

Secondly, to better understand the time and memory complexities of VI systems, additional metrics such as floating point operations per second (FLOPS) and peak memory usage can be reported. The goal then would be to use such metrics to compare entire workflows when performing matching for many queries against a large corpus, from feature extraction up through the final similarity estimation steps. Although not all VI research needs to aim for industrial-level scalability, such information can be useful for comparing application-oriented VI systems.

\subsection{Accuracy Gaps} 
Improving the accuracy of VI systems has been the main goal of most VI research, and we now highlight a number of ideas to accelerate the progress toward this goal.
The commonly used features described in Section~\ref{sec:input_repr} have been successful at capturing relevant information for quantifying similarities between versions for most mainstream music. However, in some edge cases (e.g.,~cross-genre versions, \textit{a cappella} versions, and versions of drum solos), such features may fail drastically. Therefore, to further improve the accuracy of current systems, other musical dimensions should be considered. The biggest challenge here is to find ways to exploit these uncommon musical characteristics while keeping in mind the principal invariances a VI system must consider. For example, for certain cases, a particular rhythmic pattern may facilitate identification, but using only rhythmic information for identifying versions would certainly fail in a large body of popular music where the rhythmic patterns are similar for various compositions.

A musical dimension that has not been considered in previous works is the lyrics.
Lyrics could be a major factor for improving VI accuracy for versions 
that share the same lyrics/rhyme patterns but little else,
such as those 
with no prominent melodic or harmonic characteristics to rely on,
those with greatly varying singing styles, or those with lyrics as the only connecting property.
Estimating lyrics from directly polyphonic audio is a challenging task that has been explored, but with limited success. 
However, as has been done for harmony and melody, features capturing approximate lyric information (e.g.,~phoneme-related features of the singing voice) could be a useful, complementary signal to the existing feature set.
Additionally, onset and rhythm information is not yet well-captured by the existing features, yet they are key features for determining similarity between certain types of music, such as rap and electronic music.

Track metadata, such as tags describing the high-level musical properties, can also be a powerful, under-explored signal for VI~\cite{correya2018large}. Typically, systems aim to be invariant to such high-level characteristics, but using tags such as ``vocal'' or ``instrumental'' as a way of conditioning may improve system performances, as they could inform the systems about what kind of properties they should focus on exploiting.

As for more extreme examples, there are categories of music where none of the existing or aforementioned dimensions adequately capture what makes tracks similar or not, such as sound art, soundscapes (e.g.,~rain sounds, city streets), ambient music (e.g.,~singing bowls, drones), etc. 
In these cases, the composition is closely tied to the properties of the recording itself, such as the precise sound qualities and placement of events in time.
To address such cases, music fingerprinting techniques could be applied, e.g.,~as a pre-processing step of a VI system. 

Another potential direction for improving accuracy is to fully embrace data-driven representation learning. While hand-designing features has proven to be useful for VI, it introduces a bias toward what VI systems are able to model.
Alternatively, end-to-end learning paradigms could be explored, where given a sufficient amount of data, a system learns which properties of the track are most important for the task.
In particular, these techniques give systems the potential to uncover relations beyond melody/harmony that are relevant for identifying matching versions.
These techniques have seen some success in other related domains such as speech recognition and have not yet been explored for VI.

Finally, there is an opportunity for VI systems to place more focus on post-processing operations, such as version set enhancement methods.
Such operations are generally computationally cheap and are proven to improve accuracy. However, except for a few research papers in the early 2010s, this research direction has been on standby.

\subsection{Emphasis on Subfields and  Applications}\label{sec:sub-fields}
Most VI research focuses on the general problem of identifying and retrieving versions of tracks, but there are a number of under-explored subfields and practical applications of these systems.
Certain subfields of VI focus on particular types of versions in order to address or exploit specific characteristics and challenges. For example, in versions of Western classical music~\cite{zalkow2020} (see the column ``Performances'' in Fig.~\ref{fig:version_types}), the musical variations are typically limited to those in tempo, timing, key, ``noise,'' and possibly structure (e.g.,~the presence or absence of repeats).
Therefore, the systems designed to be used in such cases focus more on being robust to noise and timing distortions while assuming 
melodic and harmonic characteristics are likely to be shared between versions.

Rather than identifying full tracks, there is the interesting subproblem of identifying versions of short queries, or phrases (e.g.,~3--15 seconds)~\cite{muller2005audio}.
Considering the difficulties that music fingerprinting systems have with identifying versions (even live performances), such an application scenario could address a particular need for end-users. 
Moreover, given the ability to identify versions of short phrases, their origins could be identified, which could enable musicologists to create phylogenetic trees of musical phrases. 
Using these, musical influences within and between musical genres and styles can be studied to have a better understanding of the evolution of musical practice.


Another less-explored application is in ``setlist identification,'' wherein the task is to identify versions from a long recording consisting of a sequence of versions of different tracks.
The long recording could be, for example, a live recording of a concert, a DJ set, or a medley.
Such long recordings are usually processed with overlapping windows that span typically 1--2 minutes. However, this is error-prone, as the windows may cross multiple tracks. To avoid this, a segmentation step can be performed beforehand, but such algorithms may also introduce erroneous segments, especially when live tracks are interrupted briefly for banter or applause. Therefore, solving problems other than identification performance may be crucial for a VI system to be used for setlist identification.

In some applications, the typical setup of having a fixed corpus does not hold, and instead, the VI problem exists in an ``online'' setting.
In this case, the goal is to build a graph of connections over time (e.g.,~as new tracks are added to a corpus) in an online fashion.
In this application, there are many open problems, such as how to avoid error propagation (e.g.,~by applying version set enhancement methods), and exploring efficient ways to perform the online steps.

Finally, applications that match audio to editorial metadata have not been well-explored.
A common example is matching a registered ``composition'' which exists purely as text metadata to a corpus of tracks.
In this context, once there is at least one track connected to a composition, standard VI techniques can be employed.
Further, it is common to match new tracks to existing ``compositions'' which already have several matching tracks, moving the problem from track-to-track matching to track-to-clique matching.

VI has come a long way in the last 20~years: from early, symbolic sequence--based approaches to recent, representation learning--based ones, a great number of techniques and ideas have been studied to approach this problem that is deeply connected to the history of musical practice. However, there is still a long way to go before we can consider the problem as solved, because, all in all, ``there is nothing that says a great song cannot be interpreted at any time in any way.''\footnote{Phil Ramone.}

\section*{Acknowledgments}
F. Yesiler is supported by the MIP-Frontiers project, the European Union’s Horizon 2020 research and innovation programme under the Marie Skłodowska-Curie grant agreement No. 765068.

\ifCLASSOPTIONcaptionsoff
  \newpage
\fi

\bibliographystyle{IEEEtran}
\bibliography{all_bib.bib}
\clearpage

\begin{IEEEbiographynophoto}{Furkan Yesiler} (furkan.yesiler@gmail.com) received his two BSc degrees summa cum laude in computer engineering and industrial engineering from Koc University, Istanbul, and his MSc degree in sound and music computing from Pompeu Fabra University, Barcelona. He is currently an Early Stage Researcher / Ph.D. candidate as a part of the MIP-Frontiers project (MSCA Grant No:765068) also at Pompeu Fabra University, 08018 Barcelona, Spain. His research focuses on machine learning for audio and music applications with a focus on contrastive and self-supervised learning methods.
\end{IEEEbiographynophoto}
\vspace{-30mm}
\begin{IEEEbiographynophoto}{Guillaume Doras} (doras@ircam.fr) received his MSc. (2016) in acoustics and signal processing applied to music at Ircam/TelecomParis, and received his Ph.D (2020) in music informatics from the Sorbonne Université, with a research focused on automatic cover detection with deep learning. He is currently working as a researcher collaborating with Ircam in Paris, where he currently works on singing voice recognition problems. His research interests are related to audio and deep learning applications. Before that, he has been leading technical projects in the telecom industry for 15 years, living and working in various countries and continents.
\end{IEEEbiographynophoto}
\vspace{-30mm}

\begin{IEEEbiographynophoto}{Rachel~M.~Bittner} (rachelbittner@spotify.com) received her Ph.D. in Music Technology and Digital Signal Processing from New York University. She is a Senior Research Scientist at Spotify, France. Bittner serves as a member of the IEEE Signal Processing Society's Audio and Acoustic Signal Processing Technical Committee. In 2015 she was awarded the Chateaubriand fellowship, and in 2018 she won the New York University Steinhardt outstanding dissertation award. Her research interests include automatic music transcription, musical source separation, machine listening, and dataset creation.
\end{IEEEbiographynophoto}
\vspace{-30mm}
\begin{IEEEbiographynophoto}{Christopher J. Tralie} (ctralie@alumni.princeton.edu) received his B.S.E. degree from Princeton University and his M.S. and Ph.D. degrees from Duke University. He was a postdoctoral researcher at Duke University and Johns Hopkins University and is now an Assistant Professor of Mathematics And Computer Science At Ursinus College, Collegeville, PA 19129, USA. He is a recipient of an NSF Graduate Research Fellowship. His research focuses on multimedia data analysis, geometric and topological data analysis, and open source academic software development.
\end{IEEEbiographynophoto}
\vspace{-30mm}
\begin{IEEEbiographynophoto}{Joan Serr\`a} (joan.serra@dolby.com) received his Ph.D. in machine learning for music from Universitat Pompeu Fabra. He was a postdoctoral researcher at IIIA-CSIC, a researcher at Telef\'onica R\&D, and is now a staff researcher with Dolby Laboratories, 08018 Barcelona, Spain. His current research focuses on neural networks for audio analysis and synthesis.
\end{IEEEbiographynophoto}

\end{document}